\documentclass[twocolumn,english,aps,superscriptaddress,pre]{revtex4-1}

\usepackage{babel}
\usepackage{amsmath}
\usepackage{amssymb}
\usepackage{graphicx}
\usepackage{ytableau}
\usepackage{physics}
\usepackage{hyperref}

\usepackage{comment}
\usepackage{stmaryrd}
\usepackage{epstopdf}
\usepackage{xcolor}
\usepackage{xcolor}
\usepackage{tikz}
\usetikzlibrary{decorations}
\usetikzlibrary{decorations.pathmorphing}
\usetikzlibrary{decorations.pathreplacing}
\usetikzlibrary{decorations.markings}
\usetikzlibrary{shapes.misc}
\usetikzlibrary{calc}

\hypersetup{
    colorlinks=true,
    linkcolor=blue,
    citecolor=blue,    
    urlcolor=blue,
}

\definecolor{mycol}{RGB}{10,55,130}

\begin{document}

\title{Numerical observation of $\mathrm{SU}(N)$ Nagaoka ferromagnetism}

\author{Thomas Botzung}
\affiliation{Laboratoire de Physique et Mod\'elisation des Milieux Condens\'es, Universit\'e Grenoble Alpes and CNRS, 25 avenue des Martyrs, 38042 Grenoble, France}

\author{Pierre Nataf}
\affiliation{Laboratoire de Physique et Mod\'elisation des Milieux Condens\'es, Universit\'e Grenoble Alpes and CNRS, 25 avenue des Martyrs, 38042 Grenoble, France}

\date{\today}
\begin{abstract} 
We provide numerical evidence of the Nagaoka’s theorem in the $\mathrm{SU}(N)$ Fermi-Hubbard model on various cluster geometries, such as the square, the honeycomb and the triangular lattices. In particular, by diagonalizing several finite-size clusters, we show that for one hole away from filling $1/N$, the itinerant ferromagnetism arises for $U$ (the positive on-site interaction) larger than $U_c$ (the value at the transition), which strongly depends on the coordination number $z$ and on $N$, the number of degenerate orbitals, that we vary from $N=2$ to $N=6$ in our simulations. We prove that $U_c$ is a non  decreasing function of $N$. In addition, we find that the lattice dependency is rooted in the kinetic energy of the hole. We find that large coordination numbers $z$ lower the value of $U_c$. Complementary, we explore the effect of long-range hopping on the appearance of itinerant ferromagnetism and demonstrate that it acts as an increased coordination number, protecting the ferromagnetic phase at small $U$. Finally, both the effects of the presence of some additional holes and of the finite size of the clusters are briefly discussed. 
\end{abstract}

\maketitle

\section{Introduction}
\label{sec: intro}

The Fermi-Hubbard model (FHM) has attracted considerable attention as an idealized model for strongly interacting electrons in a solid~\cite{Hubbard_1963, Gutzwiller_1963}. Despite its apparent simplicity, this model harbors tremendously rich physics~\cite{Arovas2022Mar}. For instance, the results obtained studying the FHM help us understand a plethora of phenomena in strongly correlated systems, including pairing mechanisms in unconventional superconductors~\cite{Scalapino2012Oct, Lee2006Jan}, the Mott metal-insulator transition~\cite{Imada1998Oct}, and diverse kind of magnetic orderings~\cite{Herrmann_1997,Mazurenko_2017,Szasz_2020}. 
Among these phenomena, the origin of itinerant ferromagnetism observed in some materials has stimulated many investigations in physical science for a long time~\cite{ Clifton1938Apr, Lieb1962Jan, Kanamori1963Sep, Hertz1976Aug, Tasaki1989Nov, Tasaki1992Sep, Millis1993Sep, Jo2009Sep, Liu2012Mar, Chen2013May,  Aron2015Jan,Sposetti2014May, Iaconis2016Apr}.

One prime example of a saturated itinerant
electron ferromagnetism appears for systems containing exactly one hole with an infinite
Hubbard repulsion $U$, phenomenon known as Nagaoka (or Thouless-Nagaoka)
ferromagnetism. In fact, it was initially demonstrated by Thouless for  some special bipartite lattices~\cite{Thouless1965Nov}, before being rigorously generalized to nonbipartite lattices by Nagaoka~\cite{Nagaoka1966}, Lieb~\cite{Lieb1989Mar}, and Tasaki~\cite{Tasaki1989Nov, Tasaki1998Apr}.

 Since then, several studies have focused on determining
if and when Nagaoka ferromagnetism emerged in various lattices and conditions  ranging from 
FHMs with finite $U$ ~\cite{Hanisch1993Jan, Hanisch1995Jan, Wurth1996Mar, Yun2021Dec} or with various physically realistic hole doppings~\cite{Doucot1989Aug, Fang1989Oct, Shastry1990Feb, Basile1990Mar, Barbieri1990Jun,Mielke_1993}, in multiple orbitals \cite{Li2014May,Bobrow_multi_2018} or with dispersionless (``flats'') bands in the spectrum~\cite{Strack1995May, Tasaki1998Apr, Vollhardt2007Jun}.

 Despite the simplicity of the definition of the FHM, it is difficult to analyze it consistently for finite $U$ and on lattices of dimensions larger than one. 
 Actually, the study of such a model represents a computational challenge which requires state-of-the-art quantum many-body numerical methods ~\cite{Schafer2021Mar, Qin2022Mar}, and is still not fully solved.
 
From an experimental point of view, quantum simulations using ultracold fermions in optical lattices \cite{Esslinger_2010,Mazurenko_2017} could help to answer open questions about the FHM.

From a theoretical point of view, inspired by analytical approaches in High energy physics, one original way to address the SU($2$) FHM has been to investigate the large limit $N$ of the $\mathrm{SU}(N)$ FHM as an asymptotic description of spins $1/2$, notably introduced in the context of high-temperature superconductors~\cite{Affleck_1988, affleck_exact_1986,Rokhsar_1990,Marder_1990}. 

Interestingly, these two latter ideas are somehow combined into the cold atomic
realization of the $\mathrm{SU}(N)$ invariant FHM on engineered optical traps~\cite{wu_exact_2003,Wu_review_2006,gorshkov_two_2010,Cazalilla_2014, capponi_phases_2016}.
In these set-ups, the Alkaline-earth cold atoms like $^{173}$Yb or $^{87}$Sr have nuclear spins which play the role of the $N$ (up to $N=10$) different colors or flavors of a set of atoms that can hop from one site to a neighboring site of the optical lattice.
 In fact, the continuous experimental progress in this field are such
that the experimentalists can fine control several physical parameters of the $\mathrm{SU}(N)$ FHM, like the filling, the interaction $U$, the hopping amplitudes, the geometry of the lattices, and the number of degenerate orbitals $N$.
~\cite{taie_su6_2012,hofrichter_direct_2016,Becker_2021,taie2020observation,Fallani_2022,pasqualetti2023equation}. 

Naturally, the question of itinerant ferromagnetism arises also in the $\mathrm{SU}(N)$ FHM \cite{Katsura2013,Bobrow2018,Liu_2019,Tamura_2021}.
In the infinite interaction $U$ limit, an important step towards an understanding has been made in~\cite{Katsura2013}, where the Nagaoka's theorem was extended to the $\mathrm{SU}(N)$ FHM for a restricted class of models satisfying the connectivity condition, which was later generalized in the light of graph theory ~\cite{Bobrow2018}. 

On the other hand, for finite $U$, several points should be addressed numerically,
to know quantitatively how the onset of Nagaoka ferromagnetism depends on the geometry of the lattices and on the number of degenerate orbitals $N$, and what is the impact of the range of the hoppings.

In this perspective, the recently developed numerical Exact Diagonalization (ED) scheme devised in \cite{botzung2023exact} for the $\mathrm{SU}(N)$ FHM, is a tool that is particularly well adapted to this purpose. In fact, such a method, implementing the full $\mathrm{SU}(N)$ symmetry  and which generalizes with semi-standard Young tableaux (ssYT) what was done for the Heisenberg $\mathrm{SU}(N)$ models with standard Young tableaux (SYT)~\cite{nataf_exact_2014,wan_exact_2017}, provides us with the eigenenergies of the model directly in each irreducible representation ({\it irrep}) of $\mathrm{SU}(N)$.
As a consequence, it does not only involve a reduction of the dimensions of the matrices to diagonalize (and thus an increase of the size of the considered clusters), but it also directly gives the relevant quantum numbers in order to determine when the ground state belongs to the fully symmetric irrep.

The remainder of the paper is organized as follows. In Sec.~\ref{sec: model}, we introduce the $\mathrm{SU}(N)$ FHM. In Sec.~\ref{subsec: method}, we provide a summary of the method employed to perform ED with the full $\mathrm{SU}(N)$ symmetry on finite-size clusters. In Sec.~\ref{subsec: Nagaoka}, we briefly remind the main results of Nagaoka's theorem and its extension to $\mathrm{SU}(N)$. In Sec.~\ref{sec: results}, we show the numerical results for the $\mathrm{SU}(N)$ FHM. Firstly, in Sec.~\ref{subsec: numerical_result_lattice_N}, we illustrate the influence of the lattice geometry on several 2D clusters and of the number of colors $N$ for $N$ up to $N=6$.  Then, in Sec.~\ref{subsec: long_range}, we analyze the situation with long-range hopping. Finally, in Sec.~\ref{subsec: finite_size}, we comment briefly about the finite-size effects and in Sec.~\ref{subsec: more holes}, we discuss the situation with more than one hole, before concluding in Sec.~\ref{sec: conclusion}.

\section{The $\mathrm{SU}(N)$ Fermi-Hubbard model}
\label{sec: model}
The $\mathrm{SU}(N)$ Fermi-Hubbard model (FHM)  can be written as
\begin{equation}
\label{eq: Hamiltonian}
H = \sum_{\langle i,j \rangle}  \Big{(}  t_{ij} E_{ij}+ \text{h.c} \Big{)} + \frac{U}{2}  \sum_{i=1}^L E_{ii}^2, 
\end{equation}
where the $t_{ij}$ are the hopping amplitudes between sites $i$ and $j$ and  $U$ is the on-site density-density interaction. The $\mathrm{SU}(N)$ invariant hopping terms read
\begin{equation}
E_{ij}=E_{ji}^{\dag}=\sum_{\sigma=1}^N c^{\dag}_{i,\sigma}c_{j,\sigma}, 
\end{equation}
where the $\sigma$ are the color (or flavors) indexes.
Note that we use the following notations: Latin letters for the site indexes and Greek letters (or capital Latin letters) for the colors.
The hopping operators satisfy the commutation relation of the $\mathrm{U}(L)$ generators ($\forall{\,1 \leq i,j,k,l \leq L}$) :
\begin{equation}
\label{commutation}
[E_{ij},E_{kl}]=\delta_{jk}E_{il}-\delta_{li}E_{kj}.
\end{equation}
In Eq. \eqref{eq: Hamiltonian}, the integer parameter $N$ is {\it hidden}: This Hamiltonian should be seen as an element of the Lie algebra of the unitary group $\mathrm{U}(L)$ \footnote{or more precisely of its universal enveloping algebra since $H$ is a polynomial of the generators of the Lie Algebra of $\mathrm{U}(L)$.}. One can define a set of  number operators, flavor-raising, and lowering operators as ($\forall {\,1 \leq \sigma, \mu \leq N}$)
\begin{equation}
F^{\sigma, \mu} = \sum_{i=1}^{L} c^{\dagger}_{i,\sigma} c_{i,\mu}.
\end{equation} 
From the commutation relation of fermions, they satisfy ($\forall {\,1 \leq \sigma, \mu, \gamma, \beta \leq N}$)
\begin{equation}
\label{commutation_F}
[F^{\sigma, \mu}, F^{\gamma, \beta}] = \delta_{\mu, \gamma} F^{\sigma, \beta}  - \delta_{\beta, \sigma} F^{\gamma, \mu}.
\end{equation}
They are dual of the $E_{ij}$ operators.
Importantly, since ($\forall { \,1 \leq \sigma, \mu \leq N}$ and $\forall {\, 1 \leq i,j \leq L}$ )
\begin{equation}
\label{commutation_F_E}
[F^{\sigma, \mu}, E_{ij}] = 0,
\end{equation}
the Hamiltonian in Eq~\eqref{eq: Hamiltonian} does not only conserve the  number of fermions in each specie (or color) $M^{\sigma}\equiv F^{\sigma, \sigma}$ , but also exhibits a global $\mathrm{U}(N)$ symmetry. Consequently,  the eigenstates of $H$ are  separated into different disconnected sectors labeled by the irreducible representations (\textit{irreps}) of $\mathrm{U}(N)$.
In that case, an efficient method has been devised to work directly in these independent sectors in Ref~\cite{botzung2023exact}. Here, we summarize the most important results, and we introduce the basic definitions needed to understand the rest of the paper.

\subsection{Review of the method}
\label{subsec: method}
 \begin{figure*}[htb]
     \includegraphics[width=1\textwidth]{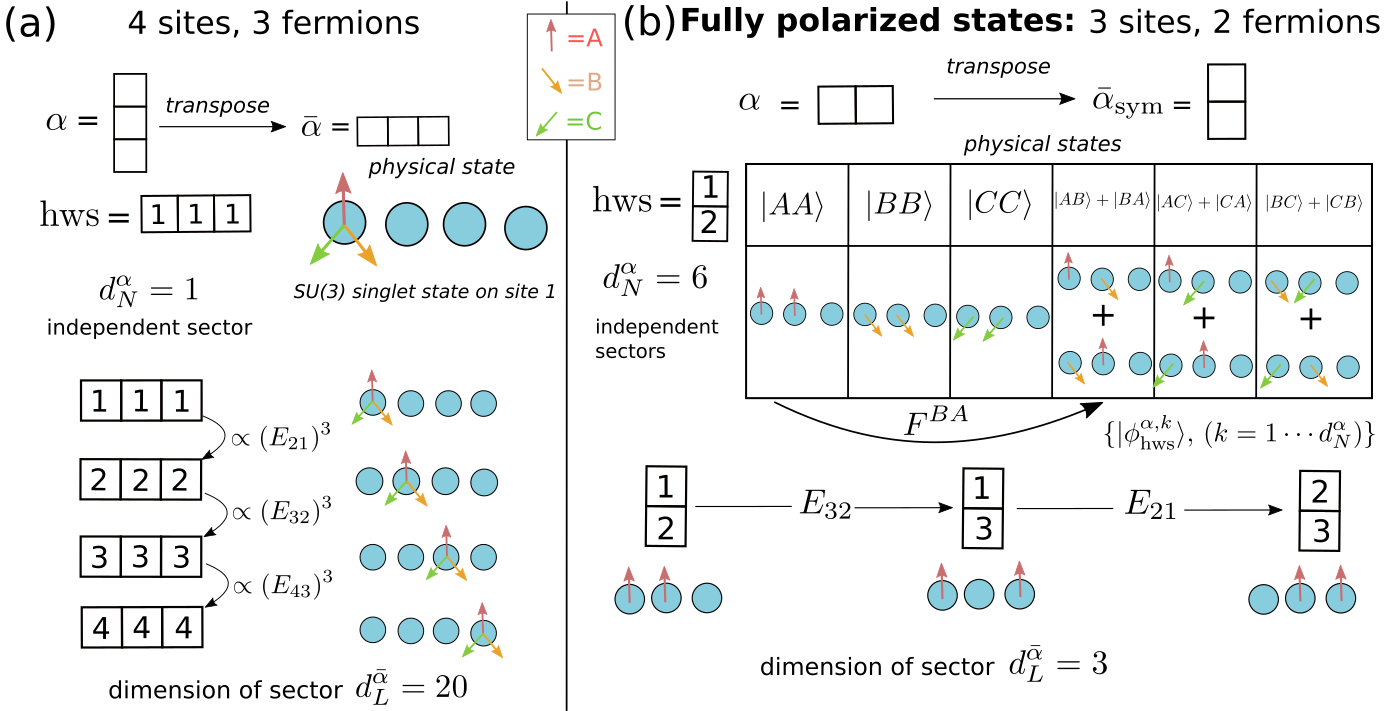}
     \caption{(a) Example of $\mathrm{SU}(3)$ irrep for $L=4$ sites  and $M=3$ fermions, labelled by tge Young diagram (YD) $\alpha=[1,1,1]$. We associate the \textit{transpose} YD, transforming rows into columns, $\bar{\alpha}=[3]$. We fill up $\bar{\alpha}$ to get the Highest Weight State (hws) to which we associate $d^{\alpha}_{N=3}$ different orthonormal physical states  which generate the $d^{\bar{\alpha}}_L$-dimensional and independent sectors $\mathcal{H}^{\bar{\alpha},k}_L$ (for $k=1 \cdots d^{\alpha}_{N=3}$, here $d^{[1,1,1]}_{N=3}=1$) under the application of the operators $E_{ij}$. (b) Examples of fully polarized states (living in the one-row irrep $\alpha=\alpha_{\text{sym}}=[M]$) for $L=3$ sites, $M=2$ fermions for $\mathrm{SU}(3)$. Similarly as (a), we present all the different physical states associated with the hws. Additionally, we show an example of the operator $F^{\sigma, \mu}$ acting on a fully polarized state, showing how it connects the different sectors $\mathcal{H}^{\bar{\alpha},k}_L$ (for $k=1 \cdots d^{\alpha}_{N=3}$, see text for details). The Young diagram representation shows the connection with the ED method. For convenience, we illustrate only one hole configuration.}
      \label{fig: method}
 \end{figure*}

In this section, we review the cornerstone principle instigating our ED procedure, i.e. the color factorization decomposition of $\mathcal{H}^{M, N}_L$, which is the Hilbert space  of $M$ SU($N$) fermions on L sites interacting through the Hamiltonian in Eq.~\eqref{eq: Hamiltonian}, and where each fermion wave-function belongs to the fundamental representation of $\mathrm{SU}(N)$. An irrep of $\mathrm{SU}(N)$ is identified by a {\it Young Diagram} (YD) or {\it shape} $\boldmath{\alpha} = [\alpha_1,\alpha_2,...,\alpha_q]$, with $q$ the number of rows of the diagram ($1 \leq q \leq N$) such that $\alpha_1\geq \alpha_2 \geq...\geq \alpha_q\geq 1 $. 
Please note that the irreps of $\mathrm{U}(N)$ or $\mathrm{SU}(N)$ are basically the same; the distinction is relevant when we include/exclude representation of generators with non vanishing trace like some combinations of $E_{ii}$ (or $F_{\sigma,\sigma})$, and when we deal with the Casimirs (cf below).
We should rather use here the common terminology, talking about $\mathrm{SU}(N)$ irreps.
In this representation, the number of particles $M=\sum_{\sigma}F^{\sigma, \sigma}=\sum_i E_{ii}$ is equal to the number of boxes, i.e. $\sum_{i=1}^N \alpha_i=M$ [cf. Fig.~\ref{fig: method} for examples]. 

It appears that $\mathcal{H}^{M, N}_L$ can be decomposed
as
\begin{equation}
\label{eq: decomposition_hilbert}
\mathcal{H}^{M, N}_L=\underset{\alpha}{\oplus}\overset{d^{\alpha}_N}{\underset{k=1}{\oplus}} \mathcal{H}^{\bar{\alpha},k}_L,
\end{equation}
where the outer sum runs over all the M-boxes YD $\alpha$ of maximum $L$ columns and $N$ rows [see Ref.~\cite{botzung2023exact} for in-depth details]. 
For a given $\alpha$, there are $d^{\alpha}_N$ independent sectors $\mathcal{H}^{\bar{\alpha},k}_L$ (for $k=1 \cdots d^{\alpha}_N$) which are invariant under the action of the Hamiltonian $H$. They are isomorphic with each other, with the same dimension $d^{\bar{\alpha}}_L$, 
where $\bar{\alpha}$ is the \textit{transpose} of a YD $\alpha$, transforming its rows into columns [cf. Fig.~\ref{fig: method}].
They will give rise to some multiple copies (some {\it multiplicities}) of the eigenspectrum corresponding to the irrep $\alpha$ in the full eigenspectrum of the Hamiltonian $H$ (cf Eq~\eqref{eq: Hamiltonian}) on $\mathcal{H}^{M, N}_L$.
In particular, the dimension $D^{M,N}_L$ of the Hilbert space $\mathcal{H}^{M,N}_L$ is given by
\begin{equation}
\label{eq: decomposition}
D^{M,N}_L\equiv \text{dim} (\mathcal{H}^{M, N}_L)=\sum_{\boldmath{\alpha}} d^{\alpha}_N  d^{\bar{\alpha}}_L,
\end{equation}

The quantity $d^{\alpha}_N $ (resp. $d^{\bar{\alpha}}_L$) stands for the dimension of the $\mathrm{SU}(N)$ irrep $\alpha$ (resp. the $\mathrm{U}(L)$ irrep $\bar{\alpha}$), that one can obtain using, e.g. the hook length formulas~\cite{Weyl1925Dec, Robinson1961}.

These dimensions are equal to the number of {\it semi-standard Young tableaux} (ssYT) of shape $\alpha$ (resp. $\bar{\alpha}$) filled with numbers from $1$ to $N$ (resp. $L$), since these latter form a basis of the $\mathrm{SU}(N)$ or $\mathrm{U}(L)$ irreps. Precisely,  a ssYT is filled up with numbers from $1$ to $N$ (resp. $L$) in non-descending order from left to right in rows and top to bottom in columns, importantly repetition is allowed in rows only.  For instance,
$$
 \ytableausetup{smalltableaux}
\raisebox{1.8ex}{$\ytableaushort{1 3 3 4, 2 4 4, 3}$}, \, \textrm{is a ssYT, while } \, \,   \ytableausetup{smalltableaux}
\raisebox{1.8ex}{$\ytableaushort{1 3 3 4, 2 4 4, 2}$},  \textrm{is not.}
$$

Moreover, for $k=1 \cdots d^{\alpha}_N$, each sector $\mathcal{H}^{\bar{\alpha},k}_L$ can be generated by the applications of the generators  $E_{ij}$  (for $1 \leq i,j \leq L$) on a state $\vert \phi_{\text{hws}}^{\alpha, k}\rangle$, which has the defining properties of the {\it Highest Weight State} (hws) of the $\mathrm{U}(L)$ irrep $\bar{\alpha}$ ~\cite{Paldus2021Jan}, 
\begin{align}
    E_{ii} \vert \phi_{\text{hws}}^{\alpha, k}\rangle &=\bar{\alpha}_i\vert \phi_{\text{hws}}^{\alpha, k}\rangle, \, \, \, \forall i \in \llbracket 1;L \rrbracket \\
    E_{ij} \vert \phi_{\text{hws}}^{\alpha, k}\rangle &=0,  \,  \, \, \textrm{for} \, i<j.
\end{align}
where $\bar{\alpha}_i$ is the number of fermions on site $i$. 

For a given $\bar{\alpha}$, the ssYT representing the hws has its first row filled with $1$, its second row filled with $2$, and so on. For example,  for $L=10$ sites, with $M=9$ particles, and for $\alpha=[5, 4]$, we have
$$
\alpha =  \ytableausetup{smalltableaux}
\raisebox{1ex}{$\ydiagram{5,4}$}  \implies \bar{\alpha} = \raisebox{1ex}{$\ydiagram{2,2, 2, 2, 1}$} \implies \text{hws} = \raisebox{1.8ex}{$\ytableaushort{1 1,  2 2, 3 3, 4 4, 5}$}, 
$$
[cf. Fig.~\ref{fig: method} for other examples]. 
In $\mathcal{H}^{M, N}_L$, for a given shape $\alpha$, there are $d^{\alpha}_N$ orthonormal states  $\vert \phi_{\text{hws}}^{\alpha, k}\rangle , (k=1 \cdots d^{\alpha}_N)$, and we pass from one to another by applying the operators $F^{\sigma, \mu}$ [cf. Fig.~\ref{fig: method}]. And from Eq.~\eqref{commutation_F_E}, we pass from one sector $\mathcal{H}^{\bar{\alpha},k}_L$ to another $\mathcal{H}^{\bar{\alpha},k'}_L$
(for $k'\neq k$) through the same operations.

Each sector $\mathcal{H}^{\bar{\alpha},k}_L$  independently represents the $\mathrm{U}(L)$ irreps $\bar{\alpha}$ (for $k=1 \cdots d^{\alpha}_N$), so that 
our algorithm is simple.
Targetting a global $\mathrm{SU}(N)$ irrep $\alpha$, we first generate the basis of ssYT of $\bar{\alpha}$, that we denote $\{ \ket{\nu} \}$.
This is a very convenient basis, where the matrix elements of the infinitesimal generators between equal or consecutive sites $E_{p p}$, $E_{p-1 p}$, $E_{p p-1}$, take simple form derived from group theory results due to Gelfand and Tsetlin \cite{Gelfand_1950}. Calling $\vert \nu \rangle$ a ssYT, one has  for $p=1\cdots L$:

\begin{align}
E_{p  p} \vert \nu \rangle &=\big{(} \# p \in \nu \big{)}\vert \nu \rangle, 
\end{align}
where $(\# p \in \nu)$ is equal to the number of occurrences of $p$ inside $\vert \nu \rangle$, corresponding thus to the occupation number on each site [see examples in Fig.~\ref{fig: method}]. 
And we have also, for $p=2\cdots L$:
\begin{align}
E_{p-1 p} \vert \nu \rangle & = \sum_{j=1}^{p-1} a^j_{p-1} \vert \nu^{+j}_{p-1}\rangle, \\
E_{p p-1} \vert \nu \rangle &=\sum_{j=1}^{p-1} b^j_{p-1} \vert \nu^{-j}_{p-1}\rangle, \\
\end{align}

where $\vert \nu^{+j}_{p-1}\rangle$ (resp. $\vert \nu^{-j}_{p-1}\rangle$)
is the same ssYT as $\vert \nu \rangle$, except that we have transformed $p$ into $p-1$ (resp. $p-1$ into $p$) in the $j^{th}$  row in $\vert \nu \rangle$. As for the coefficients  $a^j_{p-1}$ and $b^j_{p-1}$, which vanish in case such  transformations are not possible either because there is no $p$ (resp. $p-1$) in the $j^{th}$ row of $\vert \nu \rangle$,
either because the resulting tableau is not a proper ssYT, they read \cite{Vilenkin_vol3}:
\begin{align}
a^j_{p-1} & =  \left | \frac{\prod_{i=1}^p (l_{i,p}-l_{j,p-1})\prod_{i=1}^{p-2} (l_{i,p-2}-l_{j,p-1}-1)}{ \prod_{i\neq j} (l_{i,p-1}-l_{j,p-1})\prod_{i\neq j} (l_{i,p-1}-l_{j,p-1}-1)} \right | ^{1/2}, \\
b^j_{p-1} & =  \left | \frac{\prod_{i=1}^p (l_{i,p}-l_{j,p-1} + 1 )\prod_{i=1}^{p-2} (l_{i,p-2}-l_{j,p-1})}{ \prod_{i\neq j} (l_{i,p-1}-l_{j,p-1})\prod_{i\neq j} (l_{i,p-1}-l_{j,p-1}+1)} \right | ^{1/2},
\end{align}
where $l_{k,q}=m_{k,q}-k$ with $m_{k,q}$ the length of the $k^{th}$ row of the sub-tableau that remains when we delete all the boxes containing numbers $>q$ in $\vert \nu \rangle$. For instance, for the $\mathrm{SU}(4)$ adjoint irrep $\alpha = [4 3 2 0]$ for $L=10$ and $M=9$ (the basis has then $566280$ elements), we have:
 \begin{align}
\label{eq_ssYT}
 \ytableausetup{smalltableaux}
 E_{2 3}\,\raisebox{1.8ex}{$\ytableaushort{1123, 334, 55}$}=\raisebox{1.8ex}{$\ytableaushort{1122, 334, 55}$}+\sqrt{\frac{5}{3}}\,\raisebox{1.8ex}{$\ytableaushort{1123, 234, 55}$}.
\end{align}
Moreover, from successive applications of the commutation relations Eq. \eqref{commutation}, the generators $E_{p p+j} \, (j>1)$ are deduced from the generators $E_{p p+q}$ ($q<j$) through $E_{p p+j}=[E_{p p+1},E_{p+1 p+j}]$. Additionally, by using the hermitian conjugate properties of the matrices representing the operators $E_{ij}$, i.e. $E_{ij}=E_{ji}^{\dag}$, one gets the matrix representing $H$ defined in Eq.~\ref{eq: Hamiltonian} in the irrep $\bar{\alpha}$, which corresponds to the $\mathrm{SU}(N)$  irrep $\alpha$. 
The eigenvalues of the matrices are the eigenergies of $H$, with a multiplicity equal to $d^{\alpha}_N$.

 Finally, it will be useful for our purpose to characterize the irreps of U($N$) or SU($N$) by the values of some polynomial invariant operators, or Casimirs.

Among the Casimirs of U($N$) and U($L$),  the two simplest ones are the linear and the quadratic one:
\begin{align}
I_1&=\sum_i E_{ii} = \sum_{\sigma} F^{\sigma,\sigma}, \\
I_2&=\sum_{i,j} E_{ij}E_{ji} = - \sum_{\sigma,\mu} F^{\sigma,\mu}  F^{\mu, \sigma} + M(L+N),
\end{align}
which commute with all the  U($L$) generators $E_{ij}$ and with the U($N$) generators $F^{\sigma,\mu}$, as a simple consequence of the commutation rules in Eq. (\ref{commutation}) and (\ref{commutation_F}) \cite{Paldus2021Jan}.
From Schur Lemma, on a given irrep $\alpha=[\alpha_1,\alpha_2,\cdots,\alpha_N]$ of U($N$), they take constant values \cite{Paldus2021Jan},
$\chi (I_1)=\sum_i \alpha_{i}=M,$ and $
\chi (I_2)=\sum_i \alpha_{i}^2-\sum_j \bar{\alpha_{j}}^2+NM$.
Note that this is consistent with the transpose operation to pass from the U($N$) irrep $\alpha$ to the U($L$) irrep $\bar{\alpha}$.

The quadratic Casimir of SU($N$) that we call $C_2$, is a quadratic polynomial in the SU($N$) (traceless) generators, and a linear combination of $I_1$ and $I_2$ so that the vanishing commutation with the SU($N$) generators
can also be seen as a simple consequence of the properties of the U($N$) invariants.
A natural choice \cite{botzung2023exact} for $C_2$, that we use in some of the plots of the current paper is 
$C_2 =  I_2-I_1^2/N$ 
so that the constant values on an irrep $\alpha=[\alpha_1,\alpha_2,\cdots,\alpha_N]$ of SU($N$) is:
\begin{equation}
\label{formula_C_2}
\chi (C_2)=\sum_i \alpha_{i}^2-\sum_j \bar{\alpha_{j}}^2+NM-M^2/N,
\end{equation}
in agreement with \cite{pilch_formulas_1984}.

\subsection{Nagaoka's theorem}
\label{subsec: Nagaoka}

This section summarizes Nagaoka's theorem~\cite{Nagaoka1966} and its extension to $\mathrm{SU}(N)$~\cite{Katsura2013}. The approach used here has been developed in~\cite{Katsura2013, Bobrow2018}. Let us consider the subspace with a given content in colors $\{ M^{\mu}\}  = \{M^A, M^B, M^C, \cdots\}$, where the $M^{\sigma}$ are fixed and such that $M=\sum_{\sigma=1}^N M^{\sigma}=L-1$ (exactly one hole).
Such a subspace, that we can name $\mathcal{H}^{\{ M^{\mu}\}}_L$, contains many-body states that belong to different $\mathrm{SU}(N)$ irreps $\alpha$ \footnote{When $M^A=L-1$ and $M^{\sigma}=0$ for $\sigma>1$: then only the fully symmetric (one row) irrep $\alpha$ is included in the considered subspace.}, and from a simple argument based on the Perron-Frobenius theorem, one can show that the ground state of the Hamiltonian $H$ on $\mathcal{H}^{\{ M^{\mu}\}}_L$ is, under simple conditions, fully symmetric.
In the limit $U \to \infty$, every site has exactly one fermion apart from the site with a hole.
Thus, the basis  states spanning $\mathcal{H}^{\{ M^{\mu}\}}_L$ have the form
 \begin{equation}
 \label{basis_hole}
    \ket{h, \{\sigma \}} = (-1)^{h} \prod_{j\neq h, j=1}^{L} c_{j,\sigma_j }^{\dagger} \ket{0},
\end{equation}
where $ \ket{0} $ is the vacuum (no-particle) state, $\{ \sigma\}=\{ \sigma_1, \sigma_2, \cdots \sigma_L\}$ is a color configuration of content $\{ M^{\mu}\}$, i.e $\sigma_j$ is the color on site $j$ for $j=1 \cdots L$, and $h$ is the location of the hole. The $c_{j, \sigma_j}^{\dagger}$ are defined with an arbitrary (but fixed) ordered sequence on the lattice sites.  If the hopping matrix element is positive $t_{ij}>0$ ---albeit in the case of bipartite lattices this condition can be relaxed since the sign can be changed by a gauge transformation $c_{j, \sigma}^{\dagger} \to - c_{j, \sigma}^{\dagger} $ ---, the Hamiltonian satisfies a nonpositivity condition in that all its elements are 0 or $-t_{ij}$. Additionally, we suppose that the Hamiltonian satisfies the connectivity condition, which states that for any two basis elements $\ket{h, \{\sigma \}}$ and $\ket{h', \{\sigma'\}}$, there is a positive integer $n$ such that,
\begin{equation}
    \bra{h', \{\sigma' \}} H^{n} \ket{h, \{\sigma \}}\neq 0.
\end{equation}

It means that any spatial configuration of spins and a hole within $\mathcal{H}^{\{ M^{\mu}\}}_L$ can be converted to any other spatial spins and a hole configurations via a sequence of hole hoppings.
As shown in ~\cite{Bobrow2018}, one sufficient condition for the connectivity condition to hold for the  $\mathrm{SU}(N)$ FHM  is to have a non separable\footnote{Non separable means that it is still path connected if one site is withdrawn} lattice, other than the $\Theta_0$ graph (i.e a single hexagon with an additional vertex-i.e site- in the center connecting two opposite vertices) and the polygons (i.e closed chains with hoppings between nearest neighbors) with $L\geq 4$, with the additional condition for bipartite lattices that $L\geq N+2$.
Note that Nagaoka's ferromagnetism in one-dimensional chains was investigated in \cite{Xavier_2020}.

If the connectivity and the nonpositivity conditions are reunited for $H$, the Perron-Frobenius theorem is applicable~\cite{Katsura2013}, implying that the lowest-energy state in $\mathcal{H}^{\{ M^{\mu}\}}_L$ is unique and is given by a certain linear combination of all configurations $\ket{h, \{\sigma \}}$ with {\bf positive} coefficients, i.e the ground state reads
\begin{equation}
    \ket{\phi^{\{M^{\mu}\}}_{\rm{gs}}} = \sum_{h, \sigma} \alpha_{h, \{ \sigma \}} \ket{h, \{\sigma \}}, 
\end{equation}
with $ \alpha_{h, \{ \sigma \}}  > 0$. This state corresponds to a fully polarized state, i.e. living in the one row irrep $\alpha=[L-1]$. In order to see that, one can construct a trivial state with equal weight $ \ket{\phi_{\rm{t}}} = \sum_{h, \sigma} \ket{h, \{\sigma \}} $ which is by definition fully symmetric. Since we have $\bra{\phi_{\rm{t}}}\ket{\phi^{\{M^{\mu}\}}_{\rm{gs}}} > 0$,  $\ket{\phi^{\{M^{\mu}\}}_{\rm{gs}}}$ and $\ket{\phi_{\rm{t}}}$ are in the same $\mathrm{SU}(N)$ irrep (otherwise their overlap is null) and thus $\ket{\phi^{\{M^{\mu}\}}_{\rm{gs}}}$ is fully polarized. 

It is worth noticing that fully polarized states in different sectors  $\mathcal{H}^{\{ M^{\mu}\}}_L$ can be constructed from the state $\ket{\phi^{A}_{\rm{gs}}}\equiv \ket{\phi^{\{L-1,0,\cdots,0\}}_{\rm{gs}}}$, which has flavor $A$ only (i.e $M^A=L-1, M^B=M^C=...=0$), by successive application of the operators $F^{\sigma, \mu}$ (for $1 \leq \sigma, \mu \leq N$).

\section{Results and discussion}
\label{sec: results}

In this paper, we investigate the stability against finite on-site interaction $U$ and number of degenerate orbitals $N$ of Nagaoka's ferromagnetism in the $\mathrm{SU}(N)$ FHM on various lattice geometries with periodic boundary conditions. 
In particular, for our numerical numerical simulations we consider (unless otherwise specified) $M=L-1$ particles (exactly one hole), and finite-size clusters which all satisfy the connectivity condition, which is easily fulfilled for the standard 2D lattices with hopping between nearest neighbors (providing that $L\geq N+2$ for bipartite lattices).
Moreover, we choose positive hopping amplitude:  $\forall\,\langle i j \rangle: \,t_{ij}\equiv t>0\,$ (set to $1$ in our numerical simulations).
Thus, in the limit $U \to +\infty$, the existence of Nagaoka ferromagnetism is a consequence of the theorem reviewed above~\cite{Katsura2013,Bobrow2018}, and the appearance of the ferromagnetism at finite value of $U$ must be addressed numerically.
We used the method explained in Sec.~\ref{subsec: method} to perform ED with $N$ up to 6 on small clusters, such as the hexagonal, square and triangular lattices for $L=10$, $L=12$ sites, as depicted in Fig.~\ref{fig: fig1} (a), (c), (e). For larger lattice sizes (e.g. $16$ sites), the numerical analysis is restricted to $N=2$. 
To find out the ground state within the full Hilbert space for various parameters, we had to consider independently all the possible irreps so that the symmetry-resolved ED method we employed was particularly adapted.
Some basic information about the systems we address in this work (size of Hilbert space, lattice site, maximum dimension of irrep) is summarized in the Table.~\ref{table: table1}.
\begin{table}[h]
    \centering
\begin{tabular}{|c|c|c|c|} 
\hline
 $\mathrm{SU}(N)$ & $L$ & $2^{NL} \approx $ &  Max$(d^{\bar{\alpha}}_L)=$ \\
 \hline
    2   & 10 & 10 $\times 10^{5}$ & 2.7720  $\times 10^{4}$ \\
     \hline
    2   & 12 & 1.7  $\times 10^{7}$ &  3.39768 $\times 10^{5}$ \\
     \hline
    2   & 16 & 4.3 $\times 10^{9}$  &  56.632576 $\times 10^{6}$ \\
     \hline
    3   & 10 & 1.1 $\times 10^{9}$ & 3.04920  $\times 10^{5}$ \\
     \hline
    3   & 13 & 5.5  $\times 10^{11}$ &  44.660616 $\times 10^{6}$ \\
     \hline
    4   & 10 & 1.1 $\times 10^{10}$ & 5.66280  $\times 10^{5}$ \\
     \hline
    4   & 12 & 2.8 $\times 10^{14}$ &  152.252100 $\times 10^{6}$ \\
     \hline
    5   & 10 & 1.1 $\times 10^{15}$ & 8.49420 $\times 10^{5}$ \\
     \hline
    6   & 10 & 1.2  $\times 10^{18}$ &  7.50750 $\times 10^{5}$ \\
 \hline
\end{tabular}
 \caption{Table of the maximum dimension of the matrices to diagonalize , i.e Max$(d^{\bar{\alpha}}_L)$, for different lattice sizes and various $N$ for one hole (i.e $M=L-1$) compared with the dimensions of the full Hilbert space, i.e $\sum_M D^{M,N}_L=2^{NL}$.}
\label{table: table1}
\end{table}

\subsection{The effect of the number of degenerate orbitals $N$ and of the lattice structure}
\label{subsec: numerical_result_lattice_N}

\subsubsection{Spectrum analysis}
 \begin{figure}[htb]
\includegraphics[width=0.95\columnwidth]{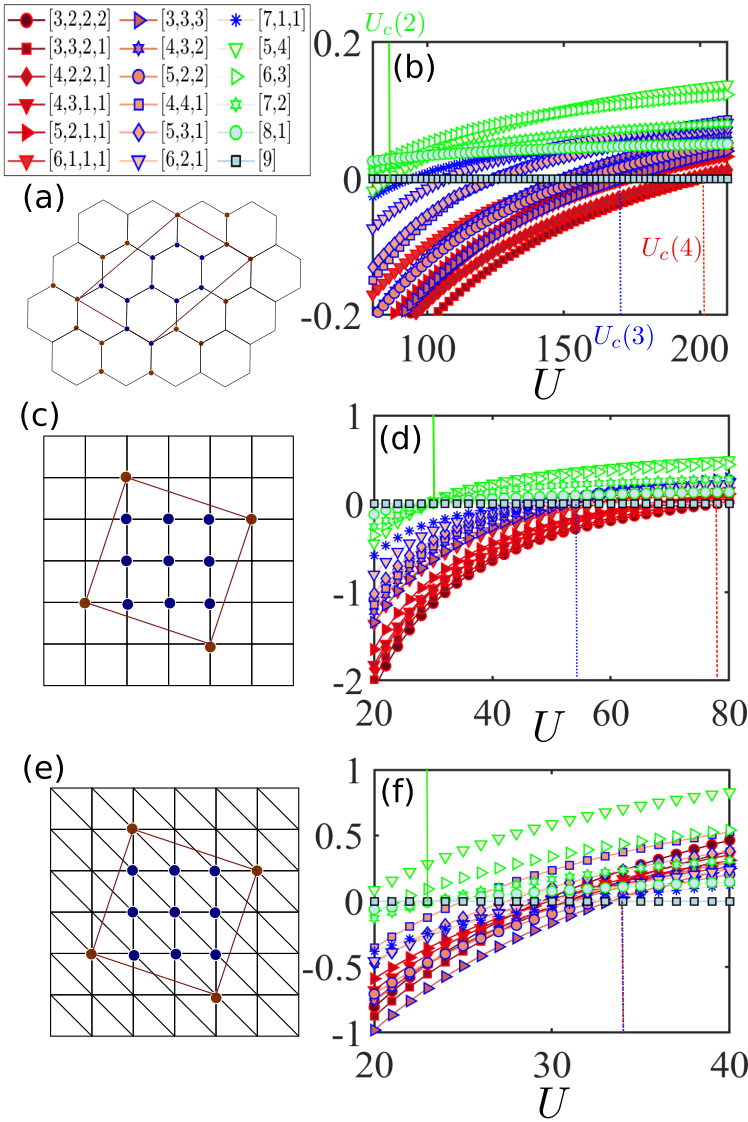}
\caption{Panels (a), (c), and (e) show sketches of 10-sites clusters of the hexagonal, the square, and the triangular lattices, respectively. The lowest energy of the SU($N$) FHM (with $t_{ij}=1 \, \forall \langle i,j \rangle$) for each irrep of $\mathrm{SU}(4)$ for $M=L-1$ fermions, for the hexagonal, the square, and the triangular lattice in panels (b), (d), and (f), respectively. To each energy, we substract the lowest energy of the fully symmetric irrep $\alpha_{\text{sym}}$, which are represented as blue-grey square dots ($=0$). Each irrep of $\mathrm{SU}(N)$ has an edge color code depending on its number of rows, red for $N=4$, blue for $N=3$, green for $N=2$. All irreps are shown in the legend in the top left.  
}
\label{fig: fig1}
\end{figure}

\begin{figure*}[htb]
\includegraphics[width=.9\textwidth]{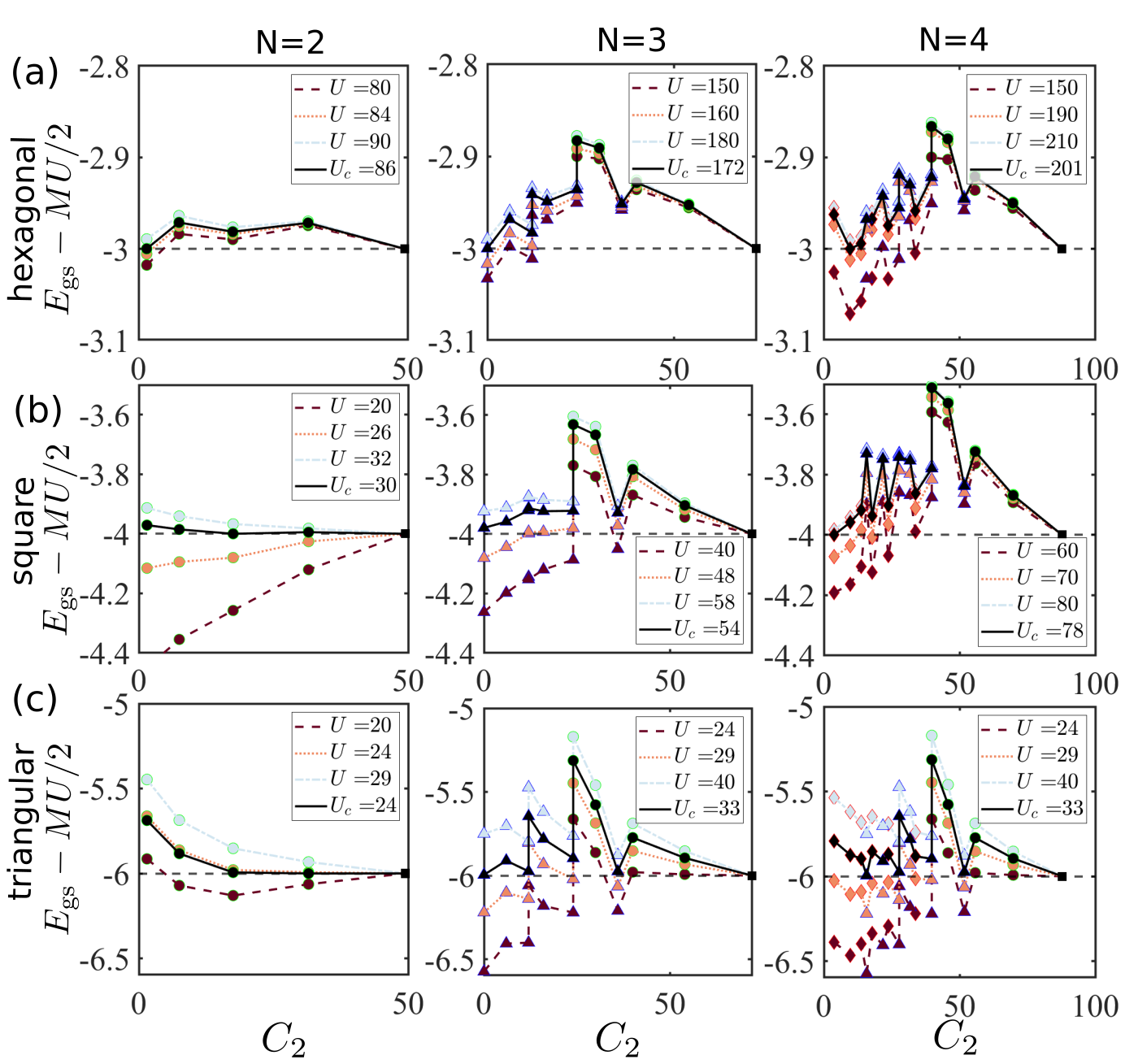}
\caption{Ground state energy $E_{\rm{gs}} - MU/2$ of the SU($N$) FHM (with $t_{ij}=t=1 \, \forall \langle i,j \rangle$) versus the quadratic Casimir $C_2$ of all the irreps of  $\mathrm{SU}(2)$, $\mathrm{SU}(3)$, and $\mathrm{SU}(4)$ from left to right for the 10-sites hexagonal (a), square (b), and triangle (c).  The lines are a guide to the eye corresponding to Eq.~\eqref{eq: E_alpha}. The critical values in the legend are rounded to the nearest integer and are depicted in black.}
\label{fig: fig2}
\end{figure*}

\begin{figure*}[htb]
\includegraphics[width=0.95\textwidth]{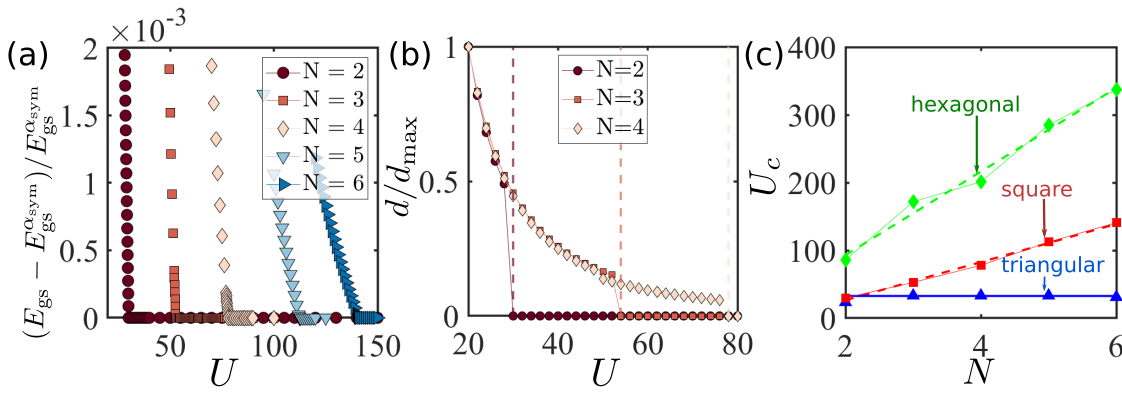}
\caption{ (a) Difference between the ground state energy of the SU($N$) FHM and the lowest energy of the fully symmetric irrep $\alpha_{\text{sym}}=[L-1]$ as a function of $U$ for the $L=10$ square lattice and various $N$. (b) Average occupation number, rescaled by its maximal value (at $U=20$) $d_{\text{max}}$, of the ground state as a function of $U$. This quantity goes to zero at the transition in agreement with (a). (c) Extracted critical $U_c$ from (a) versus $N$. The lines are best fits of the form $aN + b$. 
}
\label{fig: fig3}
\end{figure*}

In order to study the effect of the number of degenerate orbitals $N$ as well as the impact of the coordination number $z$, i.e. the number of nearest neighbors, we start by looking at the complete spectrum of various cluster geometry lattices. In Fig.~\ref{fig: fig1} (b), (d), (f), we present the lowest energies in each irrep of $\mathrm{SU}(4)$  as a function of $U$ for the hexagonal, the square, and the triangular lattice, respectively. For each energy, we withdraw the lowest energy of the irrep of fully polarized states $\alpha_{\text{sym}}=[M]$, corresponding to a M-boxes single column $\bar{\alpha}_{\text{sym}}=[1,1,..\bar{\alpha}_M=1]$. With this shift, the lowest energies of $\alpha_{\text{sym}}$ are obviously equal to zero. They are represented as blue-grey square dots with black edges in Fig.~\ref{fig: fig1} (b), (d), (f). It is worth remembering that by looking at the $\mathrm{SU}(4)$, we study as well the symmetry sector $ N < 4$, since for $N<N'$ all the irreps contained in $\mathrm{SU}(N)$ are also contained in $\mathrm{SU}(N')$. As a consequence, each irrep can be associated with a symmetry sector $N$ corresponding directly to its number of rows. To highlight the different cases, in Fig.~\ref{fig: fig1},  we choose a color code for each $N$. The irreps with $4$ rows have red edges, with $3$ rows blue edges, and with $2$ rows green edges. The respective irreps are shown in the legend.

In Fig.~\ref{fig: fig1}, we indicate by color lines ---with a color code matching the one used for the irreps of $N$--- the on-site interaction $U$ where the ground state starts to be in the irrep $\alpha_{\text{sym}}$. This provides a critical $U_c$ which locates the transition to a Nagaoka transition, for which the ground state becomes ferromagnetic. For the hexagonal [Fig.~\ref{fig: fig1} (b)] and the square lattice [Fig.~\ref{fig: fig1} (d)], we see that $U_c$ increases with $N$. For the triangular lattice [Fig.~\ref{fig: fig1} (f)], $U_c$ increases between $N=2$ and $N=3$, but it seems to become independent of $N$ for $N>3$. Indeed, in this case, we observe that the lowest energies for $U < U_c$ belongs to an irrep of  $\mathrm{SU}(3)$. 

This tendency can be understood from the properties of the $\mathrm{SU}(N)$ group. For $N < N'$, by definition, all the $\mathrm{SU}(N)$  irreps are also $\mathrm{SU}(N')$ irreps. Importantly, we note then that the most symmetric irrep is contained in all $N$. This implies that when one finds an irrep with an energy lower than the minimal one in $\alpha_{\text{sym}}$ for $\mathrm{SU}(N)$ (for a given $U \leq U_c$), it is also true for $\mathrm{SU}(N')$. As a consequence, $U_c(N) \leq U_c(N')$. The ferromagnetic can thus only appear for a higher (or equal) value of $U$ when $N$ increases.

For completeness, in Fig~\ref{fig: fig2}, we present the results of the ground state energy $E_{ \rm{gs}} - MU/2$  as a function of the quadratic Casimir $C_2$ of all the irreps of  $\mathrm{SU}(2)$, $\mathrm{SU}(3)$, $\mathrm{SU}(4)$ from left to right for the 10-sites cluster hexagonal [Fig.~\ref{fig: fig2} (a)], square [Fig.~\ref{fig: fig2} (b)], and triangle [Fig.~\ref{fig: fig2} (c)]. For clarity, the irreps of 2 rows ($N=2$) are represented by circle dots, of 3 rows ($N=3$) by triangle dots, and 4 rows ($N=4$) by diamond dots with color edges matching the previous color code. The critical values in the legend are rounded to the nearest integer and are depicted in black. 

When $N=2$, for the square [(b)] and the triangle [(c)] in Fig.~\ref{fig: fig2},  in the vicinity of the transition, the curve flattens, indicating possible near degeneracy of all spin states, as observed in larger lattices in~\cite{Sujun2021}. In contrast, this is not the case for the hexagonal lattice [(c)], where only two irreps have the same energy, reducing the degeneracy.  Above the critical value $U_c$, we see that the ground state energy is given by $E^{\alpha_{\text{sym}}}_{ \rm{gs}}-MU/2 =  -zt $ [cf. Eq.~\eqref{eq: E_alpha}] indicating a ferromagnetic phase.  In symmetry sectors of higher energies, we find a distinctively different behavior of $E_{ \rm{gs}}$. First, two different irreps can have the same Casimir $C_2$, leading to strong oscillations in the spectrum. Consequently, the curve is no longer flat near the critical point. We point out that some irreps have still the same energy at $U_c$. Lastly, we confirm that for the triangular lattice, as seen in Fig.~\ref{fig: fig3}, $U_c$ is independent of $N$ for $N \geq 3$, since the irrep with the lowest energy always belongs to $\mathrm{SU}(3)$.  

\subsubsection{Location of the transition}

Now, to locate precisely the transition, in Fig.~\ref{fig: fig3} (a), we show the difference between the ground state energy and the lowest energy of  $\alpha_{\text{sym}}$ as a function of $U$ for different $N$ [cf. color code] for the 10-sites square lattice. When this indicator nullifies, the ground state belongs to the fully symmetrized irrep, and there is thus ferromagnetism in the system. In an experimental setup, however, the complete spectrum is often non-accessible. It is easier to target appropriate observables. An interesting quantity to look at is the double occupation on sites. Since Nagaoka's ferromagnetism appears in the sector without double occupation, we naturally expect this indicator to become null in the ferromagnetic phase. We define the average occupation number as
\begin{equation}
    d = \frac{1}{L} \sum_{i=1}^{L} E_{ii}^2 - \frac{M}{L}.
\end{equation}
In Fig.~\ref{fig: fig3} (b), we show the expectation value of $\langle d \rangle$ (rescaled by its maximum value $d_{\rm{max}}$) taken in the ground state of a $10$-sites square lattice as a function of $U$. We see that at the critical $U_c$ this quantity becomes zero in complete agreement with Fig.~\ref{fig: fig3} (a). 
Fig.~\ref{fig: fig3} (c) summarizes our main results by showing $U_c$ as a function of $N$ for different lattice geometries. The dashed lines are the best fit of the forms $a N + b$ and are a guide to the eye.

\subsubsection{Influence of the lattice structure: analysis at $U=0$ and $U=+\infty$}
From Fig.~\ref{fig: fig1} and Fig.~\ref{fig: fig3}, we further find that $U_c$ strongly depends on the lattice structure. Precisely, we observe that $U_c$ decreases with an increasing coordination number $z$. For instance, for $N=3$, the critical $U_c \approx 170$ for the hexagonal lattice [$z=3$], $U_c \approx 55$ for the square [$z=4$], and $U_c \approx 34$ [$z=6$] for the triangular lattice.  We can rationalize this observation by considering the Hamiltonian in both limit $U=0$ and $U \to \infty$. 

At $U=0$, the Hamiltonian solely contains linear expression in the hopping terms and is diagonalizable in the Fourier space. It reads
\begin{equation}
\label{eq: kspace_HamiltonianP}
H =  \sum_{k_j} \epsilon(k_j) \tilde{E}_{k_jk_j},
\end{equation}
where  $\epsilon(k_j)$ is the dispersion relation: The allowed wavevectors $k_j$ are determined by the lattice, and we order them by ascending energies, i.e $\epsilon(k_i)\leq\epsilon(k_j)$ for $1\leq i < j \leq L$.
The operators $\tilde{E}_{k_jk_i}$ are the hopping terms from the mode $k_i$ to the mode $k_j$, and can be seen as some 
rotated $\mathrm{U}(L)$ generators, which also satisfy the commutation relation in Eq.~\eqref{commutation}.

We denote the eigenstates as $ \ket{n_{k}} \equiv \ket{n_{k_1} n_{k_2} ... n_{k_L}}$, where $n_{k_j}$ is the number of modes ($\leq N$) with a specific wavevector $k_j$ . 
From this representation, one can readily obtain $E^{\alpha}_{\rm{gs}}$, i.e the lowest energy in each irrep $\alpha$ by filling the associated YD of the irrep with the different modes $k$. In this scenario, the length of each row of the shape $\bar{\alpha}$ indicates the $n_{k_j}$ ($j=1,..., L$). Then, the energy for $U=0$ can be simply written as
\begin{equation}
\label{eq: Ek}
    E^{\alpha}_{\rm{gs}} (U=0) = \sum_j^{L} \bar{\alpha}_j\epsilon(k_j).
\end{equation}

\begin{figure*}[htb]
\includegraphics[width=0.9\textwidth]{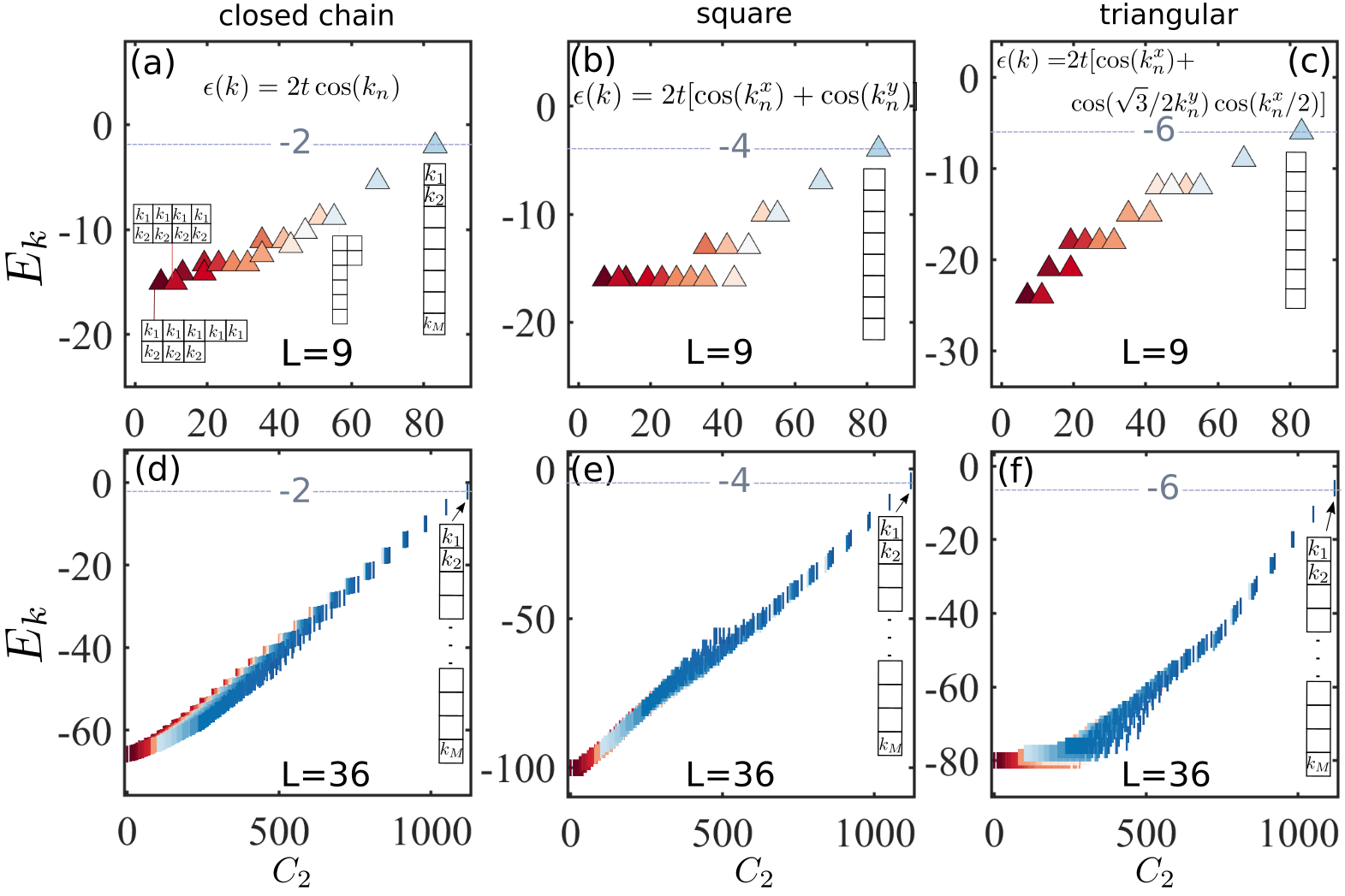}
\caption{Lowest energy of the SU($N$) FHM (with $t_{ij}=1$ and $U=0$) for each $\mathrm{SU}(5)$ irrep as a function of the Casimir $C_2$ for a closed chain (a), (d), a square lattice (b), (e), and a triangular lattice (c), (f), for $L=9$ and $L=36$ sites respectively. In (a) we illustrate the construction of Eq.~\eqref{eq: Ek} in order to find the minimal energy in each irrep. Dashed lines indicate the energy Eq.~\eqref{eq: E_alpha}.}
\label{fig: figUzero}
\end{figure*}

Fig.~\ref{fig: figUzero} shows the lowest energy of the SU($N$) FHM with $U=0$ for each $\mathrm{SU}(5)$ irrep for $L=9$  sites for a closed chain (a), a square lattice (b), and a triangular lattice (c). An example of the construction related to Eq.~\eqref{eq: Ek} is shown in (a). In Fig.~\ref{fig: figUzero}, (d), (e), (f), we show the spectrum for a larger cluster size of $36$ sites. 

It is worth noticing that for the fully symmetric irrep $\alpha_{\text{sym}}$,
the calculations are simple. 
Since $\bar{\alpha}_{\text{sym}}=[1,1,..\bar{\alpha}_M=1]$, one has $E^{\alpha_{\text{sym}}}_{\rm{gs}}=\sum_j^M\epsilon(k_j)$, which gives for $M=L-1$:
\begin{equation}
\label{eq: E_alpha}
    E^{\alpha_{\text{sym}}}_{\rm{gs}} (U=0)=-zt=E^{\alpha_{\text{sym}}}_{\rm{gs}}-MU/2.
\end{equation}
The first equality can be shown by reminding that this is equal to the lowest energy of the subspace with color $A$ only, (i.e such that $M^A=L-1, M^B=M^C=...=0$).
Then, there is no double occupancy, the $L$-dimensional basis (cf Eq.~\eqref{basis_hole}) is just in one to one correspondence with $\{h\}$, the $L$ possible locations of the hole, so that we end up with a simple tight-binding model on a regular lattice for which the ground state eigenvector $\ket{\phi^{A}_{\rm{gs}}}$ is $(1/\sqrt{L})(1,1,...1)^T$, with eigenenergy given by \eqref{eq: E_alpha}.
Secondly, since there is no double occupancy, $E_{ii}^2 \equiv E_{ii}$ ($\forall i =1 \cdots L$), and when we substract the constant $MU/2$ from the eigenergies of the FHM defined in Eq.~\eqref{eq: Hamiltonian}, they become independent of $U$. Consequently,  the limit $U \to \infty$ (or $U > 0$) and $U=0$ are completely equivalent for the fully symmetric irrep $\alpha_{\text{sym}}$.  

Now, following Nagaoka's theorem, we immediately see that the global ground state energy of the FHM, i.e $E_{\rm{gs}}$ also satisfies $E_{\rm{gs}}-MU/2=-zt$ in the limit $U\to \infty$.

\begin{figure}[t]
\includegraphics[width=0.8\columnwidth]{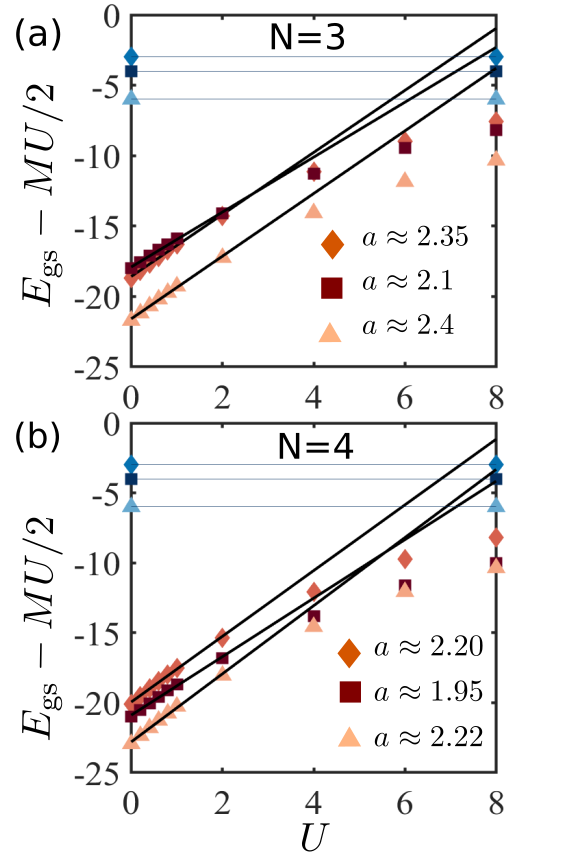}
\caption{Ground state energy  of the SU($N$) FHM $E_{\rm{gs}} - MU/2$ and that of the fully symmetric irrep $\alpha_{\text{sym}}$ equal to Eq.~\eqref{eq: E_alpha} (lines)  as a function of $U$ for the 10-sites hexagonal [diamond dots], square [square dots], and triangle [triangle dots] for $N=3$ (a), and $N=4$ (b). The black lines are best fits of the form $aU+ b$.}
\label{fig: fig4}
\end{figure}

 Qualitatively, we can now understand the $z$-dependence of $U_c$, from the behavior of the kinetic energy of the hole. In fact, since the energy of the hole is lowered when $z$ increases, it is more likely that the ground state becomes a Nagaoka state at a lower $U$. However, one must also consider how the energy of all the other irreps changes with $U$ for each lattice. An estimate can be extracted from a perturbation theory at small $U$ to determine if the perturbation is $z$-dependent. In the rotated basis, we can express the on-site interaction as 
\begin{equation}
\label{eq: perturbation_W}
    W = \frac{U}{2L} \sum_{k,l,q} \sum_{\sigma, \sigma'} \tilde{c}^{\dagger}_{k,\sigma} \tilde{c}_{k-q, \sigma} \tilde{c}^{\dagger}_{l, \sigma'} \tilde{c}_{l+q, \sigma'}.
\end{equation}
The perturbation --- in the non-degenerate case --- $\bra{n_k'}W\ket{n_k}$ depends then solely on the filling numbers $n_{k_j}$ (for $j=1 \cdots L$) and not on the coordination number. Therefore, we could expect a similar behavior at small $U$ for each of our lattices. 
In order to corroborate this statement, in Fig.~\ref{fig: fig4}, we show the ground state energy $E_{\rm{gs}} - MU/2$ as a function of $U$ for the hexagonal [diamond dots], square [square dots] and triangular [triangle dots] lattices for the two cases $N=3$ (a), and $N=4$ (b). As a guide to the eye, we present the energy $E^{\alpha_{\text{sym}}}_{\rm{gs}}-MU/2$, equal to Eq.~\eqref{eq: E_alpha}, as lines with two dots in both extremities. For $U<1$,  the ground state energy grows linearly with $U$ which appears comparable for each lattice geometry. To obtain a better estimate, we perform best fits of the form $aU+b$, depicted as black lines, and extract $a$ to demonstrate that it is undeniably lattice-independent. While we find that this is the case for $N=2$ (not shown), the situation for $N>2$ is more complex with a slope that changes slightly depending on the geometry. However, we can not attribute these differences to the coordination number. For example, in Fig.~\ref{fig: fig4} (a), the hexagonal and triangular lattice, respectively with $z=3$ and $z=6$ have the same slope. However, the square lattice ($z=4$) has a different one [cf. legend].

In this simplified picture, since $E_{\rm{gs}}$ (given by Eq.~\eqref{eq: E_alpha} at large $U$) decreases with $z$, and the energy behavior at small $U$ is comparable for all lattice, we understand that the value of $U_c$ decreases with an increasing $z$.

\begin{figure*}[tbh!]
\includegraphics[width=1\textwidth]{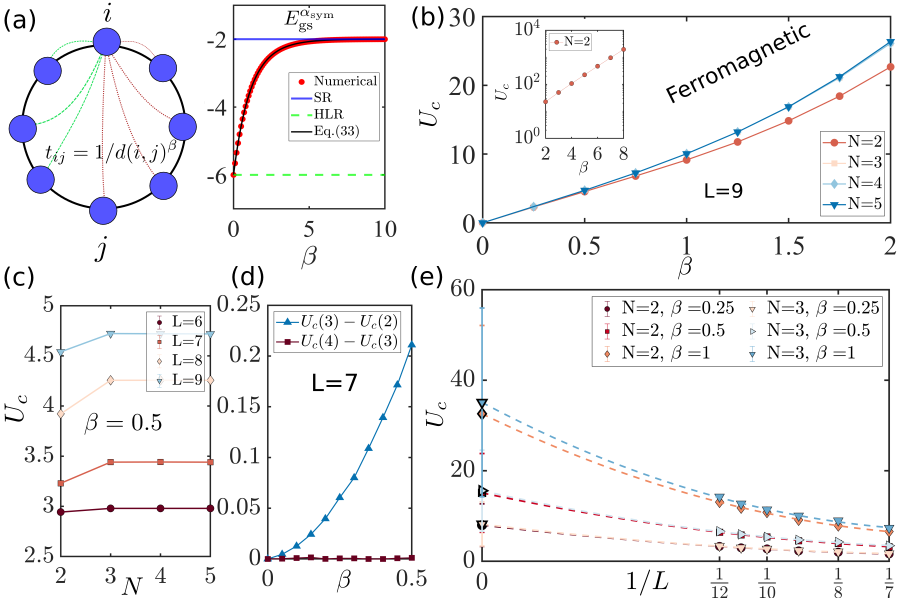}
\caption{Sketch of the chain with long-range hopping and Eq.~\eqref{eq: kinhole_lr} (black line) compared to numerical values (red dots). (b) Critical $U_c$ as a function of $\beta$ for a chain of $L=9$ sites for various $N$. For $U > U_c$ the phase is ferromagnetic. The inset shows in a semi-logarithmic scale $U_c$ at larger $\beta$ for $N=2$. (c) $U_c$ versus $N$ for different system size $L$ for $\beta=0.5$. (d) Difference between $U_c(N=2)$ and $U_c(N=3)$ (triangle dots), and difference  between $U_c(N=4)$ and $U_c(N=3)$ (square dots)  as a function of $\beta$ for $L=7$. (e) Finite-size analysis for $N=2$, $N=3$ for  $\beta=0.25$, $\beta=0.5$ and $\beta=1$. We show $U_c$ as a function of $1/L$. The dashed lines are best fits of the form $a/L^2 + b/L + c$.}
\label{fig: LR}
\end{figure*}

\subsection{The effect of the range of the hopping}
\label{subsec: long_range}
Interestingly, an extension of Nagaoka's ferromagnetism exists in the SU($2$) FHM with long-range (LR) hopping. It states, that in a homogeneous long-range (HLR) hopping model with exactly one hole,  the ground states are fully polarized states for any $U>0$~\cite{Tasaki1998Apr}. 

Is it also true for $N>2$? 
To understand the robustness of SU($N$) Nagaoka's ferromagnetism  in presence of LR hoppings, we  consider a chain of fermions with perdiodic boundary conditions and with hopping amplitudes of the form $t_{ij} = 1/d(i, j)^{\beta}$, where $d(i,j)=|i-j|$ represents the distance between sites $i$ and $j$. An example is depicted in Fig~\ref{fig: LR} (a). 
On the first hand, the case $\beta = 0$  corresponds to HLR hopping. In this case, we know from the analytical result developed in~\cite{Tasaki1998Apr} that the ferromagnetism appears at any $U>0$ for $N=2$.  
In a simplified picture, increasing the range of hopping can also be seen as increasing the coordination number, so that for general $N$, Eq.~\eqref{eq: E_alpha}
might become:
\begin{equation}
\label{eq: kinhole_lr}
E^{\alpha_{\text{sym}}}_{\rm{gs}}-MU/2= -\tilde{z}t, 
\end{equation}
where $\tilde{z} = \sum_{j=1, j\neq i_0}^{L} z/d(i_0, j)^{\beta}$, with $i_0$ an arbitrary site of reference. This hypothesis is confirmed in Fig~\ref{fig: LR} (a). Additionally, one can show that this energy is equal to the global minimal energy $E_{\rm{gs}}$ at $U=0$ (via Eq.~\eqref{eq: Ek})---which is often degenerate---. Then it appears that a finite $U$ perturbation is sufficient to increase the energy of every irrep but the fully polarized one since the latter remains unchanged by a one-site interaction perturbation. Consequently, at $\beta=0$ and for all $N$, the ground states are fully polarized for any finite $U>0$. 

On the other hand, the case $\beta \to \infty$  corresponds to a closed chain with nearest neighbors hopping, i.e. the usual short-range (SR) hopping. In this limit, we know from the Lieb-Mattis theorem that for one-dimensional systems ferromagnetism can never occur~\cite{Lieb1962}. 

Moreover, by varying the range of the hopping, i.e. $\beta$, we can study the robustness of the ferromagnetic phase. In Fig~\ref{fig: LR} (b), we show the phase diagram of the closed chain with LR hopping as a function of the exponent $\beta$ and various $N$ for $L=9$. The dots correspond to the critical $U_c$ which indicates the appearance of Nagaoka's ferromagnetism for $U>U_c$. We find that the boundary indicating the appearance of the ferromagnetic phase doesn't depend on $N$ when $N\geq3$. While in (b) the situation at low $\beta$ is unclear, we show that a difference still persists between $U_c(N=2)$ and $U_c(N\geq3)$ in Fig~\ref{fig: LR} (d) for $L=7$. Furthermore, we observe that the critical value increases rapidly with $\beta$. Inset of Fig~\ref{fig: LR} (b) shows (in a y-logarithmic scale) $U_c$ for $\beta$ up to $10$. Crucially, we see that at $\beta \to \infty$, $U_c \to \infty$, in agreement with the Lieb-Mattis theorem. These results suggest that a finite all-to-all hopping could be sufficient to observe a ferromagnetic phase at a finite $U$. To corroborate the independence of $U_c$ with $N\geq 3$, Fig~\ref{fig: LR} (c) shows $U_c$ as a function of $N$ for different sizes of the chain for $\beta=0.5$. Lastly, Fig~\ref{fig: LR} (e) shows a finite size analysis of the $U_c$ for $N=2$ and $N=3$ for $\beta=0.25, 0.5$ and $1$.We present $U_c$ as a function of $1/L$, and we perform best fits of the form $a/L^{2} + b/L +c$. We find that in the limit $L\to \infty$ the critical value is finite and increases quickly with $\beta$. However, we point out that the system sizes considered are too small to have an accurate value for the limit of $U_c$ at $L \to \infty$, at most we can give an approximate range: for instance $U_c(+\infty) \approx 16 \pm 9 $ for $N=3$ and $\beta=0.5$.

\subsection{Stability of the $\mathrm{SU}(N)$ Nagaoka states in the thermodynamic limit}

\begin{figure}[h]
\includegraphics[width=0.9\columnwidth]{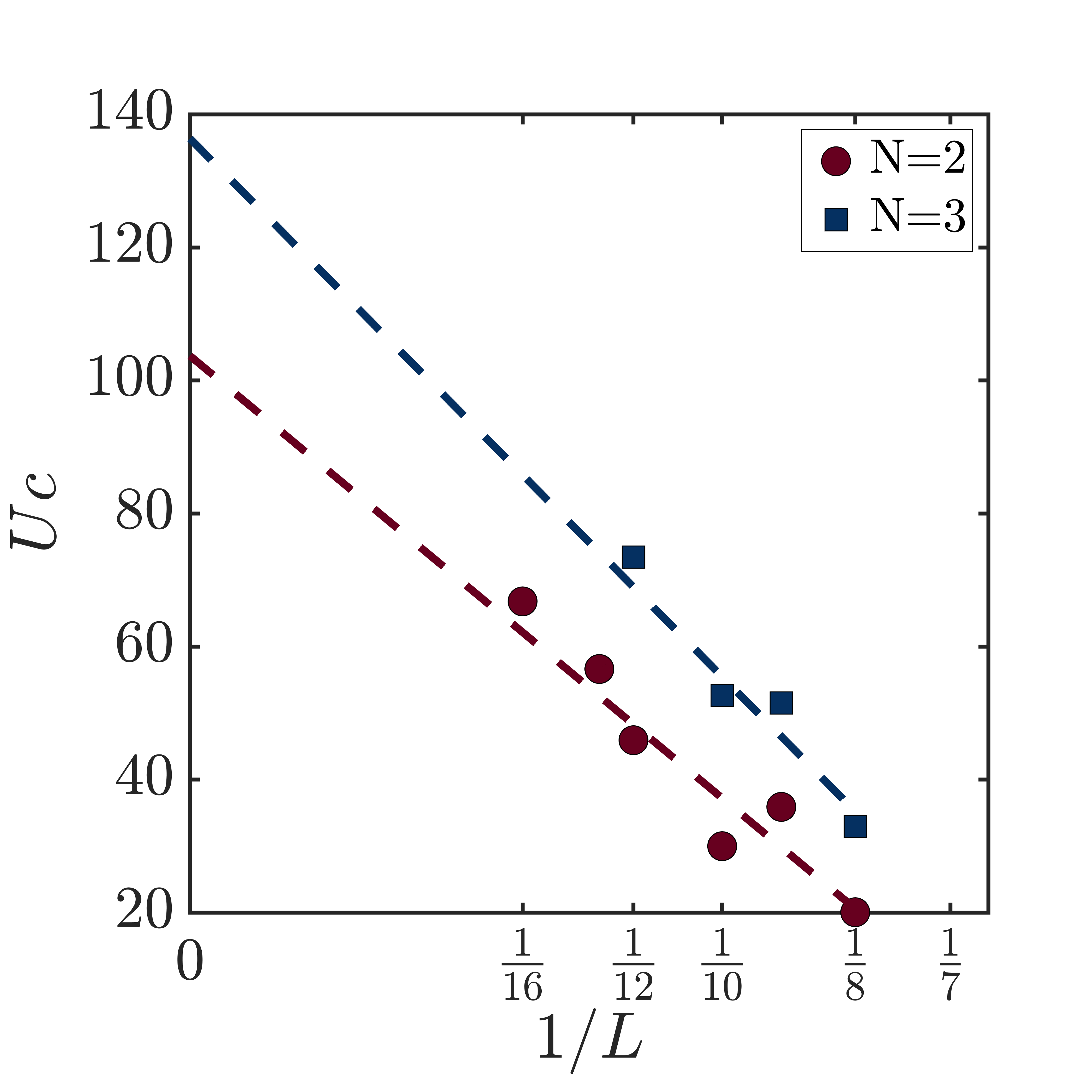}
\caption{$U_c$ versus $1/L$ for $N=2$ and $N=3$ for a square lattice with SR hopping. The lines are best fits of the form $a/L + b$. 
Note that the lattice unit vectors for the $L=12$ sites square cluster that we have used
are $t_1=(1,3)$ and $t_2=(4,0)$. For the other square clusters of size $L=8,10,9,13,16=n^2+m^2$ with $n$ and $m$ integers, they are $t_1=(n,m)$ and $t_2=(m,-n)$. }
\label{fig: scaling_size}
\end{figure}

\label{subsec: finite_size}
We comment now on the stability of the Ferromagnetic phase in the thermodynamic limit. In Fig.~\ref{fig: scaling_size}, we show the scaling of $U_c$ versus $1/L$ for $N=2$ and $N=3$ for a square lattice. We see that $U_c$ increases with the system sizes. We performed best fits of the form $a/L + b$ (lines) to highlight the general trend. From our finite-size data the critical $U_c(L\to \infty)$ remains finite : $U_c \sim 100$ for $N=2$ and $U_c \sim 140$ for $N=3$. Nevertheless, with the limited finite-size data accessible here (as well as the goodness of the fit) one cannot exclude another trend either, or even a critical $U_c \to \infty$ in the thermodynamic limit.
 For general N, addressing this difficult question clearly requires complementary numerical results based for instance on approximate schemes (such as e.g. Density Matrix Renormalization Groups (DMRG)~\cite{white_density_1992,Liu2012Mar,Sposetti2014May,Iaconis2016Apr}) which are beyond the scope of our manuscript.
Note that in the large $U$ limit, finite-temperature strong-coupling expansion is an alternative way to reveal the Nagaoka ferromagnetism of the SU($N$) FHM in the thermodynamical limit~\cite{Singh_2022}.

\subsection{Stability of the  $\mathrm{SU}(N)$ Nagaoka states against more holes}
\begin{figure}[t]
\includegraphics[width=1\columnwidth]{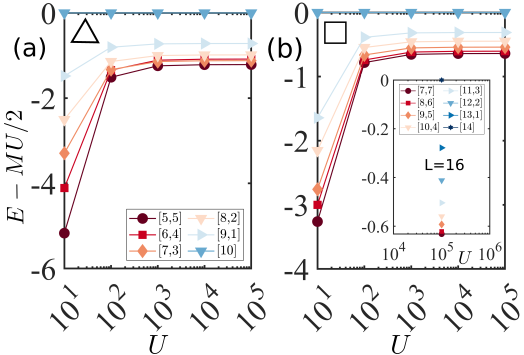}
\caption{Lowest energies (shifted by $-MU/2$) of each $\mathrm{SU}(2)$ irreps as a function of $U$ (log scale), for $L=12$ sites, on the square lattice (a) and the triangular lattice (b) for $N_h=2$ holes [cf. legend panel (a) for the irreps of $L=12$]. The behavior is similar for the hexagonal lattice (not shown). Inset shows the case $L=16$ on a square lattice for $U=10^5$.}
\label{fig: holes}
\end{figure}
\label{subsec: more holes}

Hitherto, we have only focused on the strict constraint behind Nagaoka's ferromagnetism, to have a single hole. However, a more general situation where hole density varies is also relevant. In this section, we propose to study the robustness of the ferromagnetic phase when the number of holes ($N_h)$ exceeds one.  It is convenient to note that choosing an integer number of holes is trivial with our exact diagonalization method since it amounts to having YD containing $M = L-N_h$ boxes.

For instance, we consider the situation with two holes and we restrict the study to the case $N=2$. This case is actually sufficient to conclude for all $N> 2$.
For $N=2$, we find that the ground state is never fully polarized for all the different two-dimensional lattices considered (hexagonal, square, triangular) as shown in Fig.~\ref{fig: holes}. The ground states often belong to the singlet sector, i.e. with a total spin equal to zero. The absence of ferromagnetism is immediately extendable to $N>2$, since all the $\mathrm{SU}(2)$ irreps are also $\mathrm{SU}(N)$ irreps. It means that having an irrep of $\mathrm{SU}(2)$ containing a lower energy than the energy in $\alpha_{\text{sym}}$,  remains correct for any $N> 2$. 
These results extended previous ED results obtained for $U\to \infty$~\cite{Riera1989Sep}.  Here, we have thus corroborated the absence of ferromagnetism for $N_h=2$ in several two-dimensional systems for $L\leq 12$, any $U\leq 10^5$, and any $N\geq 2$. Note that since the energies in Fig.~\ref{fig: holes} reach a plateau for large $U$, one can safely generalize the previous statement for all $U>0$. Additionally, we show (inset panel (a)) that on a square lattice of $L=16$ sites, the ferromagnetism is absent at least up to $U\leq 10^5$. 

For other densities of holes (i.e $N_h>2$), the situation can become rapidly more complex. The model for $N=2$ might accommodate ferromagnetism even in the thermodynamical limit~\cite{Riera1989Sep}.

Finally, it is worth noticing that even for a ground state living in the singlet sector, certain types of magnetism can happen. For instance, antiferromagnetism, paramagnetism, as well as low-spin coupled ferromagnetic domains could also correspond to S = 0. A recent study~\cite{Yun2023Feb} has notably highlighted the presence of ferromagnetic domains when $N_h > 1$ in the SU($2$) FHM.

\section{Conclusion}
\label{sec: conclusion}

In this work, we have studied the presence of the Ferromagnetic phase in the $\mathrm{SU}(N)$ Fermi-Hubbard model on several finite-size clusters geometries for one hole away from filling $1/N$. To do so, we have used an exact diagonalization method recently developed in~\cite{botzung2023exact}.

Firstly, we demonstrated that the appearance of the ferromagnetic phase arises for a positive on-site interaction $U$ larger than a finite $U_c$. We exhibited the fact that $U_c$ strongly depends on the coordination number $z$ and the number of degenerate orbitals $N$. While the dependency with $N$ can be directly traced back to some properties of the  $\mathrm{SU}(N)$ group, the lattice dependency can be apprehended in a free fermions hopping model framework. In this simplified picture, the kinetic energy of a hole tends to be lowered with an increasing $z$, leading to a lower value of $U_c$.

Interestingly, we extend this study to a long-range hopping framework, where the fermionic hopping can now take place between arbitrary distant lattice sites on a chain.  We motivated this picture with a hopping term suited to easily vary its range and/or strength. By doing so, we established a general picture of the robustness of Nagaoka's ferromagnetism due to $z$. 

Finally, we showed that having two holes leads to a complete loss of the ferromagnetic phase. 

Among the perspectives, one could try to address larger systems for the SU($N$) FHM to be able to extrapolate accurately the value of the critical interaction $U_c$ in the thermodynamical limit.
However difficult, such a problem could be addressed by using tensor networks/DMRG algorithms with the full SU($N$) symmetry, 
in a fashion similar to what was done for the Heisenberg SU($N$) models ~\cite{weichelbaum_nonabelian_2012,weichselbaum_unified_2018,nataf_density_2018,gozel_2020}.

 This work has been supported by an Emergence grant from CNRS Physique.

\bibliographystyle{apsrev4-1}
\bibliography{main}

\begin{thebibliography}{85}%
\makeatletter
\providecommand \@ifxundefined [1]{%
 \@ifx{#1\undefined}
}%
\providecommand \@ifnum [1]{%
 \ifnum #1\expandafter \@firstoftwo
 \else \expandafter \@secondoftwo
 \fi
}%
\providecommand \@ifx [1]{%
 \ifx #1\expandafter \@firstoftwo
 \else \expandafter \@secondoftwo
 \fi
}%
\providecommand \natexlab [1]{#1}%
\providecommand \enquote  [1]{``#1''}%
\providecommand \bibnamefont  [1]{#1}%
\providecommand \bibfnamefont [1]{#1}%
\providecommand \citenamefont [1]{#1}%
\providecommand \href@noop [0]{\@secondoftwo}%
\providecommand \href [0]{\begingroup \@sanitize@url \@href}%
\providecommand \@href[1]{\@@startlink{#1}\@@href}%
\providecommand \@@href[1]{\endgroup#1\@@endlink}%
\providecommand \@sanitize@url [0]{\catcode `\\12\catcode `\$12\catcode `\&12\catcode `\#12\catcode `\^12\catcode `\_12\catcode `\%12\relax}%
\providecommand \@@startlink[1]{}%
\providecommand \@@endlink[0]{}%
\providecommand \url  [0]{\begingroup\@sanitize@url \@url }%
\providecommand \@url [1]{\endgroup\@href {#1}{\urlprefix }}%
\providecommand \urlprefix  [0]{URL }%
\providecommand \Eprint [0]{\href }%
\providecommand \doibase [0]{http://dx.doi.org/}%
\providecommand \selectlanguage [0]{\@gobble}%
\providecommand \bibinfo  [0]{\@secondoftwo}%
\providecommand \bibfield  [0]{\@secondoftwo}%
\providecommand \translation [1]{[#1]}%
\providecommand \BibitemOpen [0]{}%
\providecommand \bibitemStop [0]{}%
\providecommand \bibitemNoStop [0]{.\EOS\space}%
\providecommand \EOS [0]{\spacefactor3000\relax}%
\providecommand \BibitemShut  [1]{\csname bibitem#1\endcsname}%
\let\auto@bib@innerbib\@empty
\bibitem [{\citenamefont {Hubbard}(1963)}]{Hubbard_1963}%
  \BibitemOpen
  \bibfield  {author} {\bibinfo {author} {\bibfnamefont {J.}~\bibnamefont {Hubbard}},\ }\href {\doibase 10.1098/rspa.1963.0204} {\bibfield  {journal} {\bibinfo  {journal} {Proceedings of the Royal Society of London. Series A. Mathematical and Physical Sciences}\ }\textbf {\bibinfo {volume} {276}},\ \bibinfo {pages} {238} (\bibinfo {year} {1963})}\BibitemShut {NoStop}%
\bibitem [{\citenamefont {Gutzwiller}(1963)}]{Gutzwiller_1963}%
  \BibitemOpen
  \bibfield  {author} {\bibinfo {author} {\bibfnamefont {M.~C.}\ \bibnamefont {Gutzwiller}},\ }\href {\doibase 10.1103/PhysRevLett.10.159} {\bibfield  {journal} {\bibinfo  {journal} {Phys. Rev. Lett.}\ }\textbf {\bibinfo {volume} {10}},\ \bibinfo {pages} {159} (\bibinfo {year} {1963})}\BibitemShut {NoStop}%
\bibitem [{\citenamefont {Arovas}\ \emph {et~al.}(2022)\citenamefont {Arovas}, \citenamefont {Berg}, \citenamefont {Kivelson},\ and\ \citenamefont {Raghu}}]{Arovas2022Mar}%
  \BibitemOpen
  \bibfield  {author} {\bibinfo {author} {\bibfnamefont {D.~P.}\ \bibnamefont {Arovas}}, \bibinfo {author} {\bibfnamefont {E.}~\bibnamefont {Berg}}, \bibinfo {author} {\bibfnamefont {S.~A.}\ \bibnamefont {Kivelson}}, \ and\ \bibinfo {author} {\bibfnamefont {S.}~\bibnamefont {Raghu}},\ }\href {\doibase 10.1146/annurev-conmatphys-031620-102024} {\bibfield  {journal} {\bibinfo  {journal} {Annu. Rev. Condens. Matter Phys.}\ }\textbf {\bibinfo {volume} {13}},\ \bibinfo {pages} {239} (\bibinfo {year} {2022})}\BibitemShut {NoStop}%
\bibitem [{\citenamefont {Scalapino}(2012)}]{Scalapino2012Oct}%
  \BibitemOpen
  \bibfield  {author} {\bibinfo {author} {\bibfnamefont {D.~J.}\ \bibnamefont {Scalapino}},\ }\href {\doibase 10.1103/RevModPhys.84.1383} {\bibfield  {journal} {\bibinfo  {journal} {Rev. Mod. Phys.}\ }\textbf {\bibinfo {volume} {84}},\ \bibinfo {pages} {1383} (\bibinfo {year} {2012})}\BibitemShut {NoStop}%
\bibitem [{\citenamefont {Lee}\ \emph {et~al.}(2006)\citenamefont {Lee}, \citenamefont {Nagaosa},\ and\ \citenamefont {Wen}}]{Lee2006Jan}%
  \BibitemOpen
  \bibfield  {author} {\bibinfo {author} {\bibfnamefont {P.~A.}\ \bibnamefont {Lee}}, \bibinfo {author} {\bibfnamefont {N.}~\bibnamefont {Nagaosa}}, \ and\ \bibinfo {author} {\bibfnamefont {X.-G.}\ \bibnamefont {Wen}},\ }\href {\doibase 10.1103/RevModPhys.78.17} {\bibfield  {journal} {\bibinfo  {journal} {Rev. Mod. Phys.}\ }\textbf {\bibinfo {volume} {78}},\ \bibinfo {pages} {17} (\bibinfo {year} {2006})}\BibitemShut {NoStop}%
\bibitem [{\citenamefont {Imada}\ \emph {et~al.}(1998)\citenamefont {Imada}, \citenamefont {Fujimori},\ and\ \citenamefont {Tokura}}]{Imada1998Oct}%
  \BibitemOpen
  \bibfield  {author} {\bibinfo {author} {\bibfnamefont {M.}~\bibnamefont {Imada}}, \bibinfo {author} {\bibfnamefont {A.}~\bibnamefont {Fujimori}}, \ and\ \bibinfo {author} {\bibfnamefont {Y.}~\bibnamefont {Tokura}},\ }\href {\doibase 10.1103/RevModPhys.70.1039} {\bibfield  {journal} {\bibinfo  {journal} {Rev. Mod. Phys.}\ }\textbf {\bibinfo {volume} {70}},\ \bibinfo {pages} {1039} (\bibinfo {year} {1998})}\BibitemShut {NoStop}%
\bibitem [{\citenamefont {Herrmann}\ and\ \citenamefont {Nolting}(1997)}]{Herrmann_1997}%
  \BibitemOpen
  \bibfield  {author} {\bibinfo {author} {\bibfnamefont {T.}~\bibnamefont {Herrmann}}\ and\ \bibinfo {author} {\bibfnamefont {W.}~\bibnamefont {Nolting}},\ }\href {\doibase https://doi.org/10.1016/S0304-8853(97)00042-5} {\bibfield  {journal} {\bibinfo  {journal} {Journal of Magnetism and Magnetic Materials}\ }\textbf {\bibinfo {volume} {170}},\ \bibinfo {pages} {253} (\bibinfo {year} {1997})}\BibitemShut {NoStop}%
\bibitem [{\citenamefont {Mazurenko}\ \emph {et~al.}(2017)\citenamefont {Mazurenko}, \citenamefont {Chiu}, \citenamefont {Ji}, \citenamefont {Parsons}, \citenamefont {Kan\'{a}sz-Nagy}, \citenamefont {Schmidt}, \citenamefont {Grusdt}, \citenamefont {Demler}, \citenamefont {Greif},\ and\ \citenamefont {Greiner}}]{Mazurenko_2017}%
  \BibitemOpen
  \bibfield  {author} {\bibinfo {author} {\bibfnamefont {A.}~\bibnamefont {Mazurenko}}, \bibinfo {author} {\bibfnamefont {C.~S.}\ \bibnamefont {Chiu}}, \bibinfo {author} {\bibfnamefont {G.}~\bibnamefont {Ji}}, \bibinfo {author} {\bibfnamefont {M.~F.}\ \bibnamefont {Parsons}}, \bibinfo {author} {\bibfnamefont {M.}~\bibnamefont {Kan\'{a}sz-Nagy}}, \bibinfo {author} {\bibfnamefont {R.}~\bibnamefont {Schmidt}}, \bibinfo {author} {\bibfnamefont {F.}~\bibnamefont {Grusdt}}, \bibinfo {author} {\bibfnamefont {E.}~\bibnamefont {Demler}}, \bibinfo {author} {\bibfnamefont {D.}~\bibnamefont {Greif}}, \ and\ \bibinfo {author} {\bibfnamefont {M.}~\bibnamefont {Greiner}},\ }\href {\doibase 10.1038/nature22362} {\bibfield  {journal} {\bibinfo  {journal} {Nature}\ }\textbf {\bibinfo {volume} {545}},\ \bibinfo {pages} {462} (\bibinfo {year} {2017})}\BibitemShut {NoStop}%
\bibitem [{\citenamefont {Szasz}\ \emph {et~al.}(2020)\citenamefont {Szasz}, \citenamefont {Motruk}, \citenamefont {Zaletel},\ and\ \citenamefont {Moore}}]{Szasz_2020}%
  \BibitemOpen
  \bibfield  {author} {\bibinfo {author} {\bibfnamefont {A.}~\bibnamefont {Szasz}}, \bibinfo {author} {\bibfnamefont {J.}~\bibnamefont {Motruk}}, \bibinfo {author} {\bibfnamefont {M.~P.}\ \bibnamefont {Zaletel}}, \ and\ \bibinfo {author} {\bibfnamefont {J.~E.}\ \bibnamefont {Moore}},\ }\href {\doibase 10.1103/PhysRevX.10.021042} {\bibfield  {journal} {\bibinfo  {journal} {Phys. Rev. X}\ }\textbf {\bibinfo {volume} {10}},\ \bibinfo {pages} {021042} (\bibinfo {year} {2020})}\BibitemShut {NoStop}%
\bibitem [{\citenamefont {Clifton}(1938)}]{Clifton1938Apr}%
  \BibitemOpen
  \bibfield  {author} {\bibinfo {author} {\bibfnamefont {S.~E.}\ \bibnamefont {Clifton}},\ }\href {\doibase 10.1098/rspa.1938.0066} {\bibfield  {journal} {\bibinfo  {journal} {Proc. R. Soc. Lond. A.}\ }\textbf {\bibinfo {volume} {165}},\ \bibinfo {pages} {372} (\bibinfo {year} {1938})}\BibitemShut {NoStop}%
\bibitem [{\citenamefont {Lieb}\ and\ \citenamefont {Mattis}(1962{\natexlab{a}})}]{Lieb1962Jan}%
  \BibitemOpen
  \bibfield  {author} {\bibinfo {author} {\bibfnamefont {E.}~\bibnamefont {Lieb}}\ and\ \bibinfo {author} {\bibfnamefont {D.}~\bibnamefont {Mattis}},\ }\href {\doibase 10.1103/PhysRev.125.164} {\bibfield  {journal} {\bibinfo  {journal} {Phys. Rev.}\ }\textbf {\bibinfo {volume} {125}},\ \bibinfo {pages} {164} (\bibinfo {year} {1962}{\natexlab{a}})}\BibitemShut {NoStop}%
\bibitem [{\citenamefont {Kanamori}(1963)}]{Kanamori1963Sep}%
  \BibitemOpen
  \bibfield  {author} {\bibinfo {author} {\bibfnamefont {J.}~\bibnamefont {Kanamori}},\ }\href {\doibase 10.1143/PTP.30.275} {\bibfield  {journal} {\bibinfo  {journal} {Prog. Theor. Phys.}\ }\textbf {\bibinfo {volume} {30}},\ \bibinfo {pages} {275} (\bibinfo {year} {1963})}\BibitemShut {NoStop}%
\bibitem [{\citenamefont {Hertz}(1976)}]{Hertz1976Aug}%
  \BibitemOpen
  \bibfield  {author} {\bibinfo {author} {\bibfnamefont {J.~A.}\ \bibnamefont {Hertz}},\ }\href {\doibase 10.1103/PhysRevB.14.1165} {\bibfield  {journal} {\bibinfo  {journal} {Phys. Rev. B}\ }\textbf {\bibinfo {volume} {14}},\ \bibinfo {pages} {1165} (\bibinfo {year} {1976})}\BibitemShut {NoStop}%
\bibitem [{\citenamefont {Tasaki}(1989)}]{Tasaki1989Nov}%
  \BibitemOpen
  \bibfield  {author} {\bibinfo {author} {\bibfnamefont {H.}~\bibnamefont {Tasaki}},\ }\href {\doibase 10.1103/PhysRevB.40.9192} {\bibfield  {journal} {\bibinfo  {journal} {Phys. Rev. B}\ }\textbf {\bibinfo {volume} {40}},\ \bibinfo {pages} {9192} (\bibinfo {year} {1989})}\BibitemShut {NoStop}%
\bibitem [{\citenamefont {Tasaki}(1992)}]{Tasaki1992Sep}%
  \BibitemOpen
  \bibfield  {author} {\bibinfo {author} {\bibfnamefont {H.}~\bibnamefont {Tasaki}},\ }\href {\doibase 10.1103/PhysRevLett.69.1608} {\bibfield  {journal} {\bibinfo  {journal} {Phys. Rev. Lett.}\ }\textbf {\bibinfo {volume} {69}},\ \bibinfo {pages} {1608} (\bibinfo {year} {1992})}\BibitemShut {NoStop}%
\bibitem [{\citenamefont {Millis}(1993)}]{Millis1993Sep}%
  \BibitemOpen
  \bibfield  {author} {\bibinfo {author} {\bibfnamefont {A.~J.}\ \bibnamefont {Millis}},\ }\href {\doibase 10.1103/PhysRevB.48.7183} {\bibfield  {journal} {\bibinfo  {journal} {Phys. Rev. B}\ }\textbf {\bibinfo {volume} {48}},\ \bibinfo {pages} {7183} (\bibinfo {year} {1993})}\BibitemShut {NoStop}%
\bibitem [{\citenamefont {Jo}\ \emph {et~al.}(2009)\citenamefont {Jo}, \citenamefont {Lee}, \citenamefont {Choi}, \citenamefont {Christensen}, \citenamefont {Kim}, \citenamefont {Thywissen}, \citenamefont {Pritchard},\ and\ \citenamefont {Ketterle}}]{Jo2009Sep}%
  \BibitemOpen
  \bibfield  {author} {\bibinfo {author} {\bibfnamefont {G.-B.}\ \bibnamefont {Jo}}, \bibinfo {author} {\bibfnamefont {Y.-R.}\ \bibnamefont {Lee}}, \bibinfo {author} {\bibfnamefont {J.-H.}\ \bibnamefont {Choi}}, \bibinfo {author} {\bibfnamefont {C.~A.}\ \bibnamefont {Christensen}}, \bibinfo {author} {\bibfnamefont {T.~H.}\ \bibnamefont {Kim}}, \bibinfo {author} {\bibfnamefont {J.~H.}\ \bibnamefont {Thywissen}}, \bibinfo {author} {\bibfnamefont {D.~E.}\ \bibnamefont {Pritchard}}, \ and\ \bibinfo {author} {\bibfnamefont {W.}~\bibnamefont {Ketterle}},\ }\href {\doibase 10.1126/science.1177112} {\bibfield  {journal} {\bibinfo  {journal} {Science}\ }\textbf {\bibinfo {volume} {325}},\ \bibinfo {pages} {1521} (\bibinfo {year} {2009})}\BibitemShut {NoStop}%
\bibitem [{\citenamefont {Liu}\ \emph {et~al.}(2012)\citenamefont {Liu}, \citenamefont {Yao}, \citenamefont {Berg}, \citenamefont {White},\ and\ \citenamefont {Kivelson}}]{Liu2012Mar}%
  \BibitemOpen
  \bibfield  {author} {\bibinfo {author} {\bibfnamefont {L.}~\bibnamefont {Liu}}, \bibinfo {author} {\bibfnamefont {H.}~\bibnamefont {Yao}}, \bibinfo {author} {\bibfnamefont {E.}~\bibnamefont {Berg}}, \bibinfo {author} {\bibfnamefont {S.~R.}\ \bibnamefont {White}}, \ and\ \bibinfo {author} {\bibfnamefont {S.~A.}\ \bibnamefont {Kivelson}},\ }\href {\doibase 10.1103/PhysRevLett.108.126406} {\bibfield  {journal} {\bibinfo  {journal} {Phys. Rev. Lett.}\ }\textbf {\bibinfo {volume} {108}},\ \bibinfo {pages} {126406} (\bibinfo {year} {2012})}\BibitemShut {NoStop}%
\bibitem [{\citenamefont {Chen}\ and\ \citenamefont {Balents}(2013)}]{Chen2013May}%
  \BibitemOpen
  \bibfield  {author} {\bibinfo {author} {\bibfnamefont {G.}~\bibnamefont {Chen}}\ and\ \bibinfo {author} {\bibfnamefont {L.}~\bibnamefont {Balents}},\ }\href {\doibase 10.1103/PhysRevLett.110.206401} {\bibfield  {journal} {\bibinfo  {journal} {Phys. Rev. Lett.}\ }\textbf {\bibinfo {volume} {110}},\ \bibinfo {pages} {206401} (\bibinfo {year} {2013})}\BibitemShut {NoStop}%
\bibitem [{\citenamefont {Aron}\ and\ \citenamefont {Kotliar}(2015)}]{Aron2015Jan}%
  \BibitemOpen
  \bibfield  {author} {\bibinfo {author} {\bibfnamefont {C.}~\bibnamefont {Aron}}\ and\ \bibinfo {author} {\bibfnamefont {G.}~\bibnamefont {Kotliar}},\ }\href {\doibase 10.1103/PhysRevB.91.041110} {\bibfield  {journal} {\bibinfo  {journal} {Phys. Rev. B}\ }\textbf {\bibinfo {volume} {91}},\ \bibinfo {pages} {041110} (\bibinfo {year} {2015})}\BibitemShut {NoStop}%
\bibitem [{\citenamefont {Sposetti}\ \emph {et~al.}(2014)\citenamefont {Sposetti}, \citenamefont {Bravo}, \citenamefont {Trumper}, \citenamefont {Gazza},\ and\ \citenamefont {Manuel}}]{Sposetti2014May}%
  \BibitemOpen
  \bibfield  {author} {\bibinfo {author} {\bibfnamefont {C.~N.}\ \bibnamefont {Sposetti}}, \bibinfo {author} {\bibfnamefont {B.}~\bibnamefont {Bravo}}, \bibinfo {author} {\bibfnamefont {A.~E.}\ \bibnamefont {Trumper}}, \bibinfo {author} {\bibfnamefont {C.~J.}\ \bibnamefont {Gazza}}, \ and\ \bibinfo {author} {\bibfnamefont {L.~O.}\ \bibnamefont {Manuel}},\ }\href {\doibase 10.1103/PhysRevLett.112.187204} {\bibfield  {journal} {\bibinfo  {journal} {Phys. Rev. Lett.}\ }\textbf {\bibinfo {volume} {112}},\ \bibinfo {pages} {187204} (\bibinfo {year} {2014})}\BibitemShut {NoStop}%
\bibitem [{\citenamefont {Iaconis}\ \emph {et~al.}(2016)\citenamefont {Iaconis}, \citenamefont {Ishizuka}, \citenamefont {Sheng},\ and\ \citenamefont {Balents}}]{Iaconis2016Apr}%
  \BibitemOpen
  \bibfield  {author} {\bibinfo {author} {\bibfnamefont {J.}~\bibnamefont {Iaconis}}, \bibinfo {author} {\bibfnamefont {H.}~\bibnamefont {Ishizuka}}, \bibinfo {author} {\bibfnamefont {D.~N.}\ \bibnamefont {Sheng}}, \ and\ \bibinfo {author} {\bibfnamefont {L.}~\bibnamefont {Balents}},\ }\href {\doibase 10.1103/PhysRevB.93.155144} {\bibfield  {journal} {\bibinfo  {journal} {Phys. Rev. B}\ }\textbf {\bibinfo {volume} {93}},\ \bibinfo {pages} {155144} (\bibinfo {year} {2016})}\BibitemShut {NoStop}%
\bibitem [{\citenamefont {Thouless}(1965)}]{Thouless1965Nov}%
  \BibitemOpen
  \bibfield  {author} {\bibinfo {author} {\bibfnamefont {D.~J.}\ \bibnamefont {Thouless}},\ }\href {\doibase 10.1088/0370-1328/86/5/301} {\bibfield  {journal} {\bibinfo  {journal} {Proc. Phys. Soc.}\ }\textbf {\bibinfo {volume} {86}},\ \bibinfo {pages} {893} (\bibinfo {year} {1965})}\BibitemShut {NoStop}%
\bibitem [{\citenamefont {Nagaoka}(1966)}]{Nagaoka1966}%
  \BibitemOpen
  \bibfield  {author} {\bibinfo {author} {\bibfnamefont {Y.}~\bibnamefont {Nagaoka}},\ }\href {\doibase 10.1103/PhysRev.147.392} {\bibfield  {journal} {\bibinfo  {journal} {Phys. Rev.}\ }\textbf {\bibinfo {volume} {147}},\ \bibinfo {pages} {392} (\bibinfo {year} {1966})}\BibitemShut {NoStop}%
\bibitem [{\citenamefont {Lieb}(1989)}]{Lieb1989Mar}%
  \BibitemOpen
  \bibfield  {author} {\bibinfo {author} {\bibfnamefont {E.~H.}\ \bibnamefont {Lieb}},\ }\href {\doibase 10.1103/PhysRevLett.62.1201} {\bibfield  {journal} {\bibinfo  {journal} {Phys. Rev. Lett.}\ }\textbf {\bibinfo {volume} {62}},\ \bibinfo {pages} {1201} (\bibinfo {year} {1989})}\BibitemShut {NoStop}%
\bibitem [{\citenamefont {Tasaki}(1998)}]{Tasaki1998Apr}%
  \BibitemOpen
  \bibfield  {author} {\bibinfo {author} {\bibfnamefont {H.}~\bibnamefont {Tasaki}},\ }\href {\doibase 10.1143/PTP.99.489} {\bibfield  {journal} {\bibinfo  {journal} {Prog. Theor. Phys.}\ }\textbf {\bibinfo {volume} {99}},\ \bibinfo {pages} {489} (\bibinfo {year} {1998})}\BibitemShut {NoStop}%
\bibitem [{\citenamefont {Hanisch}\ and\ \citenamefont {M{\ifmmode\ddot{u}\else\"{u}\fi}ller-Hartmann}(1993)}]{Hanisch1993Jan}%
  \BibitemOpen
  \bibfield  {author} {\bibinfo {author} {\bibfnamefont {{\relax Th}.}~\bibnamefont {Hanisch}}\ and\ \bibinfo {author} {\bibfnamefont {E.}~\bibnamefont {M{\ifmmode\ddot{u}\else\"{u}\fi}ller-Hartmann}},\ }\href {\doibase 10.1002/andp.19935050407} {\bibfield  {journal} {\bibinfo  {journal} {Ann. Phys.}\ }\textbf {\bibinfo {volume} {505}},\ \bibinfo {pages} {381} (\bibinfo {year} {1993})}\BibitemShut {NoStop}%
\bibitem [{\citenamefont {Hanisch}\ \emph {et~al.}(1995)\citenamefont {Hanisch}, \citenamefont {Kleine}, \citenamefont {Ritzl},\ and\ \citenamefont {M{\ifmmode\ddot{u}\else\"{u}\fi}ller-Hartmann}}]{Hanisch1995Jan}%
  \BibitemOpen
  \bibfield  {author} {\bibinfo {author} {\bibfnamefont {T.}~\bibnamefont {Hanisch}}, \bibinfo {author} {\bibfnamefont {B.}~\bibnamefont {Kleine}}, \bibinfo {author} {\bibfnamefont {A.}~\bibnamefont {Ritzl}}, \ and\ \bibinfo {author} {\bibfnamefont {E.}~\bibnamefont {M{\ifmmode\ddot{u}\else\"{u}\fi}ller-Hartmann}},\ }\href {\doibase 10.1002/andp.19955070405} {\bibfield  {journal} {\bibinfo  {journal} {Ann. Phys.}\ }\textbf {\bibinfo {volume} {507}},\ \bibinfo {pages} {303} (\bibinfo {year} {1995})}\BibitemShut {NoStop}%
\bibitem [{\citenamefont {Wurth}\ \emph {et~al.}(1996)\citenamefont {Wurth}, \citenamefont {Uhrig},\ and\ \citenamefont {M{\ifmmode\ddot{u}\else\"{u}\fi}ller-Hartmann}}]{Wurth1996Mar}%
  \BibitemOpen
  \bibfield  {author} {\bibinfo {author} {\bibfnamefont {P.}~\bibnamefont {Wurth}}, \bibinfo {author} {\bibfnamefont {G.}~\bibnamefont {Uhrig}}, \ and\ \bibinfo {author} {\bibfnamefont {E.}~\bibnamefont {M{\ifmmode\ddot{u}\else\"{u}\fi}ller-Hartmann}},\ }\href {\doibase 10.1002/andp.2065080204} {\bibfield  {journal} {\bibinfo  {journal} {Ann. Phys.}\ }\textbf {\bibinfo {volume} {508}},\ \bibinfo {pages} {148} (\bibinfo {year} {1996})}\BibitemShut {NoStop}%
\bibitem [{\citenamefont {Yun}\ \emph {et~al.}(2021{\natexlab{a}})\citenamefont {Yun}, \citenamefont {Dobrautz}, \citenamefont {Luo},\ and\ \citenamefont {Alavi}}]{Yun2021Dec}%
  \BibitemOpen
  \bibfield  {author} {\bibinfo {author} {\bibfnamefont {S.}~\bibnamefont {Yun}}, \bibinfo {author} {\bibfnamefont {W.}~\bibnamefont {Dobrautz}}, \bibinfo {author} {\bibfnamefont {H.}~\bibnamefont {Luo}}, \ and\ \bibinfo {author} {\bibfnamefont {A.}~\bibnamefont {Alavi}},\ }\href {\doibase 10.1103/PhysRevB.104.235102} {\bibfield  {journal} {\bibinfo  {journal} {Phys. Rev. B}\ }\textbf {\bibinfo {volume} {104}},\ \bibinfo {pages} {235102} (\bibinfo {year} {2021}{\natexlab{a}})}\BibitemShut {NoStop}%
\bibitem [{\citenamefont {Doucot}\ and\ \citenamefont {Wen}(1989)}]{Doucot1989Aug}%
  \BibitemOpen
  \bibfield  {author} {\bibinfo {author} {\bibfnamefont {B.}~\bibnamefont {Doucot}}\ and\ \bibinfo {author} {\bibfnamefont {X.~G.}\ \bibnamefont {Wen}},\ }\href {\doibase 10.1103/PhysRevB.40.2719} {\bibfield  {journal} {\bibinfo  {journal} {Phys. Rev. B}\ }\textbf {\bibinfo {volume} {40}},\ \bibinfo {pages} {2719} (\bibinfo {year} {1989})}\BibitemShut {NoStop}%
\bibitem [{\citenamefont {Fang}\ \emph {et~al.}(1989)\citenamefont {Fang}, \citenamefont {Ruckenstein}, \citenamefont {Dagotto},\ and\ \citenamefont {Schmitt-Rink}}]{Fang1989Oct}%
  \BibitemOpen
  \bibfield  {author} {\bibinfo {author} {\bibfnamefont {Y.}~\bibnamefont {Fang}}, \bibinfo {author} {\bibfnamefont {A.~E.}\ \bibnamefont {Ruckenstein}}, \bibinfo {author} {\bibfnamefont {E.}~\bibnamefont {Dagotto}}, \ and\ \bibinfo {author} {\bibfnamefont {S.}~\bibnamefont {Schmitt-Rink}},\ }\href {\doibase 10.1103/PhysRevB.40.7406} {\bibfield  {journal} {\bibinfo  {journal} {Phys. Rev. B}\ }\textbf {\bibinfo {volume} {40}},\ \bibinfo {pages} {7406} (\bibinfo {year} {1989})}\BibitemShut {NoStop}%
\bibitem [{\citenamefont {Shastry}\ \emph {et~al.}(1990)\citenamefont {Shastry}, \citenamefont {Krishnamurthy},\ and\ \citenamefont {Anderson}}]{Shastry1990Feb}%
  \BibitemOpen
  \bibfield  {author} {\bibinfo {author} {\bibfnamefont {B.~S.}\ \bibnamefont {Shastry}}, \bibinfo {author} {\bibfnamefont {H.~R.}\ \bibnamefont {Krishnamurthy}}, \ and\ \bibinfo {author} {\bibfnamefont {P.~W.}\ \bibnamefont {Anderson}},\ }\href {\doibase 10.1103/PhysRevB.41.2375} {\bibfield  {journal} {\bibinfo  {journal} {Phys. Rev. B}\ }\textbf {\bibinfo {volume} {41}},\ \bibinfo {pages} {2375} (\bibinfo {year} {1990})}\BibitemShut {NoStop}%
\bibitem [{\citenamefont {Basile}\ and\ \citenamefont {Elser}(1990)}]{Basile1990Mar}%
  \BibitemOpen
  \bibfield  {author} {\bibinfo {author} {\bibfnamefont {A.~G.}\ \bibnamefont {Basile}}\ and\ \bibinfo {author} {\bibfnamefont {V.}~\bibnamefont {Elser}},\ }\href {\doibase 10.1103/PhysRevB.41.4842} {\bibfield  {journal} {\bibinfo  {journal} {Phys. Rev. B}\ }\textbf {\bibinfo {volume} {41}},\ \bibinfo {pages} {4842} (\bibinfo {year} {1990})}\BibitemShut {NoStop}%
\bibitem [{\citenamefont {Barbieri}\ \emph {et~al.}(1990)\citenamefont {Barbieri}, \citenamefont {Riera},\ and\ \citenamefont {Young}}]{Barbieri1990Jun}%
  \BibitemOpen
  \bibfield  {author} {\bibinfo {author} {\bibfnamefont {A.}~\bibnamefont {Barbieri}}, \bibinfo {author} {\bibfnamefont {J.~A.}\ \bibnamefont {Riera}}, \ and\ \bibinfo {author} {\bibfnamefont {A.~P.}\ \bibnamefont {Young}},\ }\href {\doibase 10.1103/PhysRevB.41.11697} {\bibfield  {journal} {\bibinfo  {journal} {Phys. Rev. B}\ }\textbf {\bibinfo {volume} {41}},\ \bibinfo {pages} {11697} (\bibinfo {year} {1990})}\BibitemShut {NoStop}%
\bibitem [{\citenamefont {Mielke}\ and\ \citenamefont {Tasaki}(1993)}]{Mielke_1993}%
  \BibitemOpen
  \bibfield  {author} {\bibinfo {author} {\bibfnamefont {A.}~\bibnamefont {Mielke}}\ and\ \bibinfo {author} {\bibfnamefont {H.}~\bibnamefont {Tasaki}},\ }\href@noop {} {\bibfield  {journal} {\bibinfo  {journal} {Communications in Mathematical Physics}\ }\textbf {\bibinfo {volume} {158}},\ \bibinfo {pages} {341 } (\bibinfo {year} {1993})}\BibitemShut {NoStop}%
\bibitem [{\citenamefont {Li}\ \emph {et~al.}(2014)\citenamefont {Li}, \citenamefont {Lieb},\ and\ \citenamefont {Wu}}]{Li2014May}%
  \BibitemOpen
  \bibfield  {author} {\bibinfo {author} {\bibfnamefont {Y.}~\bibnamefont {Li}}, \bibinfo {author} {\bibfnamefont {E.~H.}\ \bibnamefont {Lieb}}, \ and\ \bibinfo {author} {\bibfnamefont {C.}~\bibnamefont {Wu}},\ }\href {\doibase 10.1103/PhysRevLett.112.217201} {\bibfield  {journal} {\bibinfo  {journal} {Phys. Rev. Lett.}\ }\textbf {\bibinfo {volume} {112}},\ \bibinfo {pages} {217201} (\bibinfo {year} {2014})}\BibitemShut {NoStop}%
\bibitem [{\citenamefont {Bobrow}\ and\ \citenamefont {Li}(2018)}]{Bobrow_multi_2018}%
  \BibitemOpen
  \bibfield  {author} {\bibinfo {author} {\bibfnamefont {E.}~\bibnamefont {Bobrow}}\ and\ \bibinfo {author} {\bibfnamefont {Y.}~\bibnamefont {Li}},\ }\href {\doibase 10.1103/PhysRevB.97.155132} {\bibfield  {journal} {\bibinfo  {journal} {Phys. Rev. B}\ }\textbf {\bibinfo {volume} {97}},\ \bibinfo {pages} {155132} (\bibinfo {year} {2018})}\BibitemShut {NoStop}%
\bibitem [{\citenamefont {Strack}\ and\ \citenamefont {Vollhardt}(1995)}]{Strack1995May}%
  \BibitemOpen
  \bibfield  {author} {\bibinfo {author} {\bibfnamefont {R.}~\bibnamefont {Strack}}\ and\ \bibinfo {author} {\bibfnamefont {D.}~\bibnamefont {Vollhardt}},\ }\href {\doibase 10.1007/BF00752314} {\bibfield  {journal} {\bibinfo  {journal} {J. Low Temp. Phys.}\ }\textbf {\bibinfo {volume} {99}},\ \bibinfo {pages} {385} (\bibinfo {year} {1995})}\BibitemShut {NoStop}%
\bibitem [{\citenamefont {Vollhardt}\ \emph {et~al.}(2007)\citenamefont {Vollhardt}, \citenamefont {Bl{\ifmmode\ddot{u}\else\"{u}\fi}mer}, \citenamefont {Held}, \citenamefont {Kollar}, \citenamefont {Schlipf}, \citenamefont {Ulmke},\ and\ \citenamefont {Wahle}}]{Vollhardt2007Jun}%
  \BibitemOpen
  \bibfield  {author} {\bibinfo {author} {\bibfnamefont {D.}~\bibnamefont {Vollhardt}}, \bibinfo {author} {\bibfnamefont {N.}~\bibnamefont {Bl{\ifmmode\ddot{u}\else\"{u}\fi}mer}}, \bibinfo {author} {\bibfnamefont {K.}~\bibnamefont {Held}}, \bibinfo {author} {\bibfnamefont {M.}~\bibnamefont {Kollar}}, \bibinfo {author} {\bibfnamefont {J.}~\bibnamefont {Schlipf}}, \bibinfo {author} {\bibfnamefont {M.}~\bibnamefont {Ulmke}}, \ and\ \bibinfo {author} {\bibfnamefont {J.}~\bibnamefont {Wahle}},\ }in\ \href {\doibase 10.1007/BFb0107631} {\emph {\bibinfo {booktitle} {{Advances in Solid State Physics 38}}}}\ (\bibinfo  {publisher} {Springer},\ \bibinfo {address} {Berlin, Germany},\ \bibinfo {year} {2007})\ pp.\ \bibinfo {pages} {383--396}\BibitemShut {NoStop}%
\bibitem [{\citenamefont {Sch{\ifmmode\ddot{a}\else\"{a}\fi}fer}\ \emph {et~al.}(2021)\citenamefont {Sch{\ifmmode\ddot{a}\else\"{a}\fi}fer}, \citenamefont {Wentzell}, \citenamefont {{\ifmmode\check{S}\else\v{S}\fi}imkovic}, \citenamefont {He}, \citenamefont {Hille}, \citenamefont {Klett}, \citenamefont {Eckhardt}, \citenamefont {Arzhang}, \citenamefont {Harkov}, \citenamefont {Le~R{\ifmmode\acute{e}\else\'{e}\fi}gent}, \citenamefont {Kirsch}, \citenamefont {Wang}, \citenamefont {Kim}, \citenamefont {Kozik}, \citenamefont {Stepanov}, \citenamefont {Kauch}, \citenamefont {Andergassen}, \citenamefont {Hansmann}, \citenamefont {Rohe}, \citenamefont {Vilk}, \citenamefont {LeBlanc}, \citenamefont {Zhang}, \citenamefont {Tremblay}, \citenamefont {Ferrero}, \citenamefont {Parcollet},\ and\ \citenamefont {Georges}}]{Schafer2021Mar}%
  \BibitemOpen
  \bibfield  {author} {\bibinfo {author} {\bibfnamefont {T.}~\bibnamefont {Sch{\ifmmode\ddot{a}\else\"{a}\fi}fer}}, \bibinfo {author} {\bibfnamefont {N.}~\bibnamefont {Wentzell}}, \bibinfo {author} {\bibfnamefont {F.}~\bibnamefont {{\ifmmode\check{S}\else\v{S}\fi}imkovic}}, \bibinfo {author} {\bibfnamefont {Y.-Y.}\ \bibnamefont {He}}, \bibinfo {author} {\bibfnamefont {C.}~\bibnamefont {Hille}}, \bibinfo {author} {\bibfnamefont {M.}~\bibnamefont {Klett}}, \bibinfo {author} {\bibfnamefont {C.~J.}\ \bibnamefont {Eckhardt}}, \bibinfo {author} {\bibfnamefont {B.}~\bibnamefont {Arzhang}}, \bibinfo {author} {\bibfnamefont {V.}~\bibnamefont {Harkov}}, \bibinfo {author} {\bibfnamefont {F.-M.}\ \bibnamefont {Le~R{\ifmmode\acute{e}\else\'{e}\fi}gent}}, \bibinfo {author} {\bibfnamefont {A.}~\bibnamefont {Kirsch}}, \bibinfo {author} {\bibfnamefont {Y.}~\bibnamefont {Wang}}, \bibinfo {author} {\bibfnamefont {A.~J.}\ \bibnamefont {Kim}}, \bibinfo {author} {\bibfnamefont {E.}~\bibnamefont {Kozik}}, \bibinfo {author}
  {\bibfnamefont {E.~A.}\ \bibnamefont {Stepanov}}, \bibinfo {author} {\bibfnamefont {A.}~\bibnamefont {Kauch}}, \bibinfo {author} {\bibfnamefont {S.}~\bibnamefont {Andergassen}}, \bibinfo {author} {\bibfnamefont {P.}~\bibnamefont {Hansmann}}, \bibinfo {author} {\bibfnamefont {D.}~\bibnamefont {Rohe}}, \bibinfo {author} {\bibfnamefont {Y.~M.}\ \bibnamefont {Vilk}}, \bibinfo {author} {\bibfnamefont {J.~P.~F.}\ \bibnamefont {LeBlanc}}, \bibinfo {author} {\bibfnamefont {S.}~\bibnamefont {Zhang}}, \bibinfo {author} {\bibfnamefont {A.-M.~S.}\ \bibnamefont {Tremblay}}, \bibinfo {author} {\bibfnamefont {M.}~\bibnamefont {Ferrero}}, \bibinfo {author} {\bibfnamefont {O.}~\bibnamefont {Parcollet}}, \ and\ \bibinfo {author} {\bibfnamefont {A.}~\bibnamefont {Georges}},\ }\href {\doibase 10.1103/PhysRevX.11.011058} {\bibfield  {journal} {\bibinfo  {journal} {Phys. Rev. X}\ }\textbf {\bibinfo {volume} {11}},\ \bibinfo {pages} {011058} (\bibinfo {year} {2021})}\BibitemShut {NoStop}%
\bibitem [{\citenamefont {Qin}\ \emph {et~al.}(2022)\citenamefont {Qin}, \citenamefont {Sch{\ifmmode\ddot{a}\else\"{a}\fi}fer}, \citenamefont {Andergassen}, \citenamefont {Corboz},\ and\ \citenamefont {Gull}}]{Qin2022Mar}%
  \BibitemOpen
  \bibfield  {author} {\bibinfo {author} {\bibfnamefont {M.}~\bibnamefont {Qin}}, \bibinfo {author} {\bibfnamefont {T.}~\bibnamefont {Sch{\ifmmode\ddot{a}\else\"{a}\fi}fer}}, \bibinfo {author} {\bibfnamefont {S.}~\bibnamefont {Andergassen}}, \bibinfo {author} {\bibfnamefont {P.}~\bibnamefont {Corboz}}, \ and\ \bibinfo {author} {\bibfnamefont {E.}~\bibnamefont {Gull}},\ }\href {\doibase 10.1146/annurev-conmatphys-090921-033948} {\bibfield  {journal} {\bibinfo  {journal} {Annu. Rev. Condens. Matter Phys.}\ }\textbf {\bibinfo {volume} {13}},\ \bibinfo {pages} {275} (\bibinfo {year} {2022})}\BibitemShut {NoStop}%
\bibitem [{\citenamefont {Esslinger}(2010)}]{Esslinger_2010}%
  \BibitemOpen
  \bibfield  {author} {\bibinfo {author} {\bibfnamefont {T.}~\bibnamefont {Esslinger}},\ }\href {\doibase 10.1146/annurev-conmatphys-070909-104059} {\bibfield  {journal} {\bibinfo  {journal} {Annual Review of Condensed Matter Physics}\ }\textbf {\bibinfo {volume} {1}},\ \bibinfo {pages} {129} (\bibinfo {year} {2010})},\ \Eprint {http://arxiv.org/abs/1007.0012} {arXiv:1007.0012 [cond-mat.quant-gas]} \BibitemShut {NoStop}%
\bibitem [{\citenamefont {Affleck}\ and\ \citenamefont {Marston}(1988)}]{Affleck_1988}%
  \BibitemOpen
  \bibfield  {author} {\bibinfo {author} {\bibfnamefont {I.}~\bibnamefont {Affleck}}\ and\ \bibinfo {author} {\bibfnamefont {J.~B.}\ \bibnamefont {Marston}},\ }\href {\doibase 10.1103/PhysRevB.37.3774} {\bibfield  {journal} {\bibinfo  {journal} {Phys. Rev. B}\ }\textbf {\bibinfo {volume} {37}},\ \bibinfo {pages} {3774} (\bibinfo {year} {1988})}\BibitemShut {NoStop}%
\bibitem [{\citenamefont {Affleck}(1986)}]{affleck_exact_1986}%
  \BibitemOpen
  \bibfield  {author} {\bibinfo {author} {\bibfnamefont {I.}~\bibnamefont {Affleck}},\ }\href {http://www.sciencedirect.com/science/article/pii/0550321386901677} {\bibfield  {journal} {\bibinfo  {journal} {Nuclear Physics B}\ }\textbf {\bibinfo {volume} {265}},\ \bibinfo {pages} {409 } (\bibinfo {year} {1986})}\BibitemShut {NoStop}%
\bibitem [{\citenamefont {Rokhsar}(1990)}]{Rokhsar_1990}%
  \BibitemOpen
  \bibfield  {author} {\bibinfo {author} {\bibfnamefont {D.~S.}\ \bibnamefont {Rokhsar}},\ }\href {\doibase 10.1103/PhysRevB.42.2526} {\bibfield  {journal} {\bibinfo  {journal} {Phys. Rev. B}\ }\textbf {\bibinfo {volume} {42}},\ \bibinfo {pages} {2526} (\bibinfo {year} {1990})}\BibitemShut {NoStop}%
\bibitem [{\citenamefont {Marder}\ \emph {et~al.}(1990)\citenamefont {Marder}, \citenamefont {Papanicolaou},\ and\ \citenamefont {Psaltakis}}]{Marder_1990}%
  \BibitemOpen
  \bibfield  {author} {\bibinfo {author} {\bibfnamefont {M.}~\bibnamefont {Marder}}, \bibinfo {author} {\bibfnamefont {N.}~\bibnamefont {Papanicolaou}}, \ and\ \bibinfo {author} {\bibfnamefont {G.~C.}\ \bibnamefont {Psaltakis}},\ }\href {\doibase 10.1103/PhysRevB.41.6920} {\bibfield  {journal} {\bibinfo  {journal} {Phys. Rev. B}\ }\textbf {\bibinfo {volume} {41}},\ \bibinfo {pages} {6920} (\bibinfo {year} {1990})}\BibitemShut {NoStop}%
\bibitem [{\citenamefont {Wu}\ \emph {et~al.}(2003)\citenamefont {Wu}, \citenamefont {Hu},\ and\ \citenamefont {Zhang}}]{wu_exact_2003}%
  \BibitemOpen
  \bibfield  {author} {\bibinfo {author} {\bibfnamefont {C.}~\bibnamefont {Wu}}, \bibinfo {author} {\bibfnamefont {J.-p.}\ \bibnamefont {Hu}}, \ and\ \bibinfo {author} {\bibfnamefont {S.-c.}\ \bibnamefont {Zhang}},\ }\href {https://link.aps.org/doi/10.1103/PhysRevLett.91.186402} {\bibfield  {journal} {\bibinfo  {journal} {Phys. Rev. Lett.}\ }\textbf {\bibinfo {volume} {91}},\ \bibinfo {pages} {186402} (\bibinfo {year} {2003})}\BibitemShut {NoStop}%
\bibitem [{\citenamefont {Wu}(2006)}]{Wu_review_2006}%
  \BibitemOpen
  \bibfield  {author} {\bibinfo {author} {\bibfnamefont {C.}~\bibnamefont {Wu}},\ }\href {\doibase 10.1142/S0217984906012213} {\bibfield  {journal} {\bibinfo  {journal} {Modern Physics Letters B}\ }\textbf {\bibinfo {volume} {20}},\ \bibinfo {pages} {1707} (\bibinfo {year} {2006})}\BibitemShut {NoStop}%
\bibitem [{\citenamefont {Gorshkov}\ \emph {et~al.}(2010)\citenamefont {Gorshkov}, \citenamefont {Hermele}, \citenamefont {Gurarie}, \citenamefont {Xu}, \citenamefont {Julienne}, \citenamefont {Ye}, \citenamefont {Zoller}, \citenamefont {Demler}, \citenamefont {Lukin},\ and\ \citenamefont {Rey}}]{gorshkov_two_2010}%
  \BibitemOpen
  \bibfield  {author} {\bibinfo {author} {\bibfnamefont {A.~V.}\ \bibnamefont {Gorshkov}}, \bibinfo {author} {\bibfnamefont {M.}~\bibnamefont {Hermele}}, \bibinfo {author} {\bibfnamefont {V.}~\bibnamefont {Gurarie}}, \bibinfo {author} {\bibfnamefont {C.}~\bibnamefont {Xu}}, \bibinfo {author} {\bibfnamefont {P.~S.}\ \bibnamefont {Julienne}}, \bibinfo {author} {\bibfnamefont {J.}~\bibnamefont {Ye}}, \bibinfo {author} {\bibfnamefont {P.}~\bibnamefont {Zoller}}, \bibinfo {author} {\bibfnamefont {E.}~\bibnamefont {Demler}}, \bibinfo {author} {\bibfnamefont {M.~D.}\ \bibnamefont {Lukin}}, \ and\ \bibinfo {author} {\bibfnamefont {A.}~\bibnamefont {Rey}},\ }\href {https://www.nature.com/articles/nphys1535} {\bibfield  {journal} {\bibinfo  {journal} {Nature physics}\ }\textbf {\bibinfo {volume} {6}},\ \bibinfo {pages} {289} (\bibinfo {year} {2010})}\BibitemShut {NoStop}%
\bibitem [{\citenamefont {Cazalilla}\ and\ \citenamefont {Rey}(2014)}]{Cazalilla_2014}%
  \BibitemOpen
  \bibfield  {author} {\bibinfo {author} {\bibfnamefont {M.~A.}\ \bibnamefont {Cazalilla}}\ and\ \bibinfo {author} {\bibfnamefont {A.~M.}\ \bibnamefont {Rey}},\ }\href {\doibase 10.1088/0034-4885/77/12/124401} {\bibfield  {journal} {\bibinfo  {journal} {Reports on Progress in Physics}\ }\textbf {\bibinfo {volume} {77}},\ \bibinfo {pages} {124401} (\bibinfo {year} {2014})}\BibitemShut {NoStop}%
\bibitem [{\citenamefont {Capponi}\ \emph {et~al.}(2016)\citenamefont {Capponi}, \citenamefont {Lecheminant},\ and\ \citenamefont {Totsuka}}]{capponi_phases_2016}%
  \BibitemOpen
  \bibfield  {author} {\bibinfo {author} {\bibfnamefont {S.}~\bibnamefont {Capponi}}, \bibinfo {author} {\bibfnamefont {P.}~\bibnamefont {Lecheminant}}, \ and\ \bibinfo {author} {\bibfnamefont {K.}~\bibnamefont {Totsuka}},\ }\href {http://www.sciencedirect.com/science/article/pii/S0003491616000130} {\bibfield  {journal} {\bibinfo  {journal} {Annals of Physics}\ }\textbf {\bibinfo {volume} {367}},\ \bibinfo {pages} {50 } (\bibinfo {year} {2016})}\BibitemShut {NoStop}%
\bibitem [{\citenamefont {Taie}\ \emph {et~al.}(2012)\citenamefont {Taie}, \citenamefont {Yamazaki}, \citenamefont {Sugawa},\ and\ \citenamefont {Takahashi}}]{taie_su6_2012}%
  \BibitemOpen
  \bibfield  {author} {\bibinfo {author} {\bibfnamefont {S.}~\bibnamefont {Taie}}, \bibinfo {author} {\bibfnamefont {R.}~\bibnamefont {Yamazaki}}, \bibinfo {author} {\bibfnamefont {S.}~\bibnamefont {Sugawa}}, \ and\ \bibinfo {author} {\bibfnamefont {Y.}~\bibnamefont {Takahashi}},\ }\href {https://www.nature.com/articles/nphys2430} {\bibfield  {journal} {\bibinfo  {journal} {Nature Physics}\ }\textbf {\bibinfo {volume} {8}},\ \bibinfo {pages} {825} (\bibinfo {year} {2012})}\BibitemShut {NoStop}%
\bibitem [{\citenamefont {Hofrichter}\ \emph {et~al.}(2016)\citenamefont {Hofrichter}, \citenamefont {Riegger}, \citenamefont {Scazza}, \citenamefont {H\"ofer}, \citenamefont {Fernandes}, \citenamefont {Bloch},\ and\ \citenamefont {F\"olling}}]{hofrichter_direct_2016}%
  \BibitemOpen
  \bibfield  {author} {\bibinfo {author} {\bibfnamefont {C.}~\bibnamefont {Hofrichter}}, \bibinfo {author} {\bibfnamefont {L.}~\bibnamefont {Riegger}}, \bibinfo {author} {\bibfnamefont {F.}~\bibnamefont {Scazza}}, \bibinfo {author} {\bibfnamefont {M.}~\bibnamefont {H\"ofer}}, \bibinfo {author} {\bibfnamefont {D.~R.}\ \bibnamefont {Fernandes}}, \bibinfo {author} {\bibfnamefont {I.}~\bibnamefont {Bloch}}, \ and\ \bibinfo {author} {\bibfnamefont {S.}~\bibnamefont {F\"olling}},\ }\href {\doibase 10.1103/PhysRevX.6.021030} {\bibfield  {journal} {\bibinfo  {journal} {Phys. Rev. X}\ }\textbf {\bibinfo {volume} {6}},\ \bibinfo {pages} {021030} (\bibinfo {year} {2016})}\BibitemShut {NoStop}%
\bibitem [{\citenamefont {Abeln}\ \emph {et~al.}(2021)\citenamefont {Abeln}, \citenamefont {Sponselee}, \citenamefont {Diem}, \citenamefont {Pintul}, \citenamefont {Sengstock},\ and\ \citenamefont {Becker}}]{Becker_2021}%
  \BibitemOpen
  \bibfield  {author} {\bibinfo {author} {\bibfnamefont {B.}~\bibnamefont {Abeln}}, \bibinfo {author} {\bibfnamefont {K.}~\bibnamefont {Sponselee}}, \bibinfo {author} {\bibfnamefont {M.}~\bibnamefont {Diem}}, \bibinfo {author} {\bibfnamefont {N.}~\bibnamefont {Pintul}}, \bibinfo {author} {\bibfnamefont {K.}~\bibnamefont {Sengstock}}, \ and\ \bibinfo {author} {\bibfnamefont {C.}~\bibnamefont {Becker}},\ }\href {\doibase 10.1103/PhysRevA.103.033315} {\bibfield  {journal} {\bibinfo  {journal} {Phys. Rev. A}\ }\textbf {\bibinfo {volume} {103}},\ \bibinfo {pages} {033315} (\bibinfo {year} {2021})}\BibitemShut {NoStop}%
\bibitem [{\citenamefont {Taie}\ \emph {et~al.}(2022)\citenamefont {Taie}, \citenamefont {Ibarra-Garc{\'\i}a-Padilla}, \citenamefont {Nishizawa}, \citenamefont {Takasu}, \citenamefont {Kuno}, \citenamefont {Wei}, \citenamefont {Scalettar}, \citenamefont {Hazzard},\ and\ \citenamefont {Takahashi}}]{taie2020observation}%
  \BibitemOpen
  \bibfield  {author} {\bibinfo {author} {\bibfnamefont {S.}~\bibnamefont {Taie}}, \bibinfo {author} {\bibfnamefont {E.}~\bibnamefont {Ibarra-Garc{\'\i}a-Padilla}}, \bibinfo {author} {\bibfnamefont {N.}~\bibnamefont {Nishizawa}}, \bibinfo {author} {\bibfnamefont {Y.}~\bibnamefont {Takasu}}, \bibinfo {author} {\bibfnamefont {Y.}~\bibnamefont {Kuno}}, \bibinfo {author} {\bibfnamefont {H.-T.}\ \bibnamefont {Wei}}, \bibinfo {author} {\bibfnamefont {R.~T.}\ \bibnamefont {Scalettar}}, \bibinfo {author} {\bibfnamefont {K.~R.~A.}\ \bibnamefont {Hazzard}}, \ and\ \bibinfo {author} {\bibfnamefont {Y.}~\bibnamefont {Takahashi}},\ }\href {\doibase 10.1038/s41567-022-01725-6} {\bibfield  {journal} {\bibinfo  {journal} {Nature Physics}\ }\textbf {\bibinfo {volume} {18}},\ \bibinfo {pages} {1356} (\bibinfo {year} {2022})}\BibitemShut {NoStop}%
\bibitem [{\citenamefont {Tusi}\ \emph {et~al.}(2022)\citenamefont {Tusi}, \citenamefont {Franchi}, \citenamefont {Livi}, \citenamefont {Baumann}, \citenamefont {Benedicto~Orenes}, \citenamefont {Del~Re}, \citenamefont {Barfknecht}, \citenamefont {Zhou}, \citenamefont {Inguscio}, \citenamefont {Cappellini}, \citenamefont {Capone}, \citenamefont {Catani},\ and\ \citenamefont {Fallani}}]{Fallani_2022}%
  \BibitemOpen
  \bibfield  {author} {\bibinfo {author} {\bibfnamefont {D.}~\bibnamefont {Tusi}}, \bibinfo {author} {\bibfnamefont {L.}~\bibnamefont {Franchi}}, \bibinfo {author} {\bibfnamefont {L.~F.}\ \bibnamefont {Livi}}, \bibinfo {author} {\bibfnamefont {K.}~\bibnamefont {Baumann}}, \bibinfo {author} {\bibfnamefont {D.}~\bibnamefont {Benedicto~Orenes}}, \bibinfo {author} {\bibfnamefont {L.}~\bibnamefont {Del~Re}}, \bibinfo {author} {\bibfnamefont {R.~E.}\ \bibnamefont {Barfknecht}}, \bibinfo {author} {\bibfnamefont {T.~W.}\ \bibnamefont {Zhou}}, \bibinfo {author} {\bibfnamefont {M.}~\bibnamefont {Inguscio}}, \bibinfo {author} {\bibfnamefont {G.}~\bibnamefont {Cappellini}}, \bibinfo {author} {\bibfnamefont {M.}~\bibnamefont {Capone}}, \bibinfo {author} {\bibfnamefont {J.}~\bibnamefont {Catani}}, \ and\ \bibinfo {author} {\bibfnamefont {L.}~\bibnamefont {Fallani}},\ }\href {\doibase 10.1038/s41567-022-01726-5} {\bibfield  {journal} {\bibinfo  {journal} {Nature Physics}\ }\textbf {\bibinfo {volume} {18}},\ \bibinfo {pages}
  {1201} (\bibinfo {year} {2022})}\BibitemShut {NoStop}%
\bibitem [{\citenamefont {Pasqualetti}\ \emph {et~al.}(2023)\citenamefont {Pasqualetti}, \citenamefont {Bettermann}, \citenamefont {Oppong}, \citenamefont {Ibarra-García-Padilla}, \citenamefont {Dasgupta}, \citenamefont {Scalettar}, \citenamefont {Hazzard}, \citenamefont {Bloch},\ and\ \citenamefont {Fölling}}]{pasqualetti2023equation}%
  \BibitemOpen
  \bibfield  {author} {\bibinfo {author} {\bibfnamefont {G.}~\bibnamefont {Pasqualetti}}, \bibinfo {author} {\bibfnamefont {O.}~\bibnamefont {Bettermann}}, \bibinfo {author} {\bibfnamefont {N.~D.}\ \bibnamefont {Oppong}}, \bibinfo {author} {\bibfnamefont {E.}~\bibnamefont {Ibarra-García-Padilla}}, \bibinfo {author} {\bibfnamefont {S.}~\bibnamefont {Dasgupta}}, \bibinfo {author} {\bibfnamefont {R.~T.}\ \bibnamefont {Scalettar}}, \bibinfo {author} {\bibfnamefont {K.~R.~A.}\ \bibnamefont {Hazzard}}, \bibinfo {author} {\bibfnamefont {I.}~\bibnamefont {Bloch}}, \ and\ \bibinfo {author} {\bibfnamefont {S.}~\bibnamefont {Fölling}},\ }\href@noop {} {\  (\bibinfo {year} {2023})},\ \Eprint {http://arxiv.org/abs/2305.18967} {arXiv:2305.18967 [cond-mat.quant-gas]} \BibitemShut {NoStop}%
\bibitem [{\citenamefont {Katsura}\ and\ \citenamefont {Tanaka}(2013)}]{Katsura2013}%
  \BibitemOpen
  \bibfield  {author} {\bibinfo {author} {\bibfnamefont {H.}~\bibnamefont {Katsura}}\ and\ \bibinfo {author} {\bibfnamefont {A.}~\bibnamefont {Tanaka}},\ }\href {\doibase 10.1103/PhysRevA.87.013617} {\bibfield  {journal} {\bibinfo  {journal} {Phys. Rev. A}\ }\textbf {\bibinfo {volume} {87}},\ \bibinfo {pages} {013617} (\bibinfo {year} {2013})}\BibitemShut {NoStop}%
\bibitem [{\citenamefont {Bobrow}\ \emph {et~al.}(2018)\citenamefont {Bobrow}, \citenamefont {Stubis},\ and\ \citenamefont {Li}}]{Bobrow2018}%
  \BibitemOpen
  \bibfield  {author} {\bibinfo {author} {\bibfnamefont {E.}~\bibnamefont {Bobrow}}, \bibinfo {author} {\bibfnamefont {K.}~\bibnamefont {Stubis}}, \ and\ \bibinfo {author} {\bibfnamefont {Y.}~\bibnamefont {Li}},\ }\href {\doibase 10.1103/PhysRevB.98.180101} {\bibfield  {journal} {\bibinfo  {journal} {Phys. Rev. B}\ }\textbf {\bibinfo {volume} {98}},\ \bibinfo {pages} {180101} (\bibinfo {year} {2018})}\BibitemShut {NoStop}%
\bibitem [{\citenamefont {Liu}\ \emph {et~al.}(2019)\citenamefont {Liu}, \citenamefont {Nie},\ and\ \citenamefont {Zhang}}]{Liu_2019}%
  \BibitemOpen
  \bibfield  {author} {\bibinfo {author} {\bibfnamefont {R.}~\bibnamefont {Liu}}, \bibinfo {author} {\bibfnamefont {W.}~\bibnamefont {Nie}}, \ and\ \bibinfo {author} {\bibfnamefont {W.}~\bibnamefont {Zhang}},\ }\href {\doibase https://doi.org/10.1016/j.scib.2019.08.013} {\bibfield  {journal} {\bibinfo  {journal} {Science Bulletin}\ }\textbf {\bibinfo {volume} {64}},\ \bibinfo {pages} {1490} (\bibinfo {year} {2019})}\BibitemShut {NoStop}%
\bibitem [{\citenamefont {Tamura}\ and\ \citenamefont {Katsura}(2021)}]{Tamura_2021}%
  \BibitemOpen
  \bibfield  {author} {\bibinfo {author} {\bibfnamefont {K.}~\bibnamefont {Tamura}}\ and\ \bibinfo {author} {\bibfnamefont {H.}~\bibnamefont {Katsura}},\ }\href@noop {} {\bibfield  {journal} {\bibinfo  {journal} {Journal of Statistical Physics}\ ,\ \bibinfo {pages} {16}} (\bibinfo {year} {2021})}\BibitemShut {NoStop}%
\bibitem [{\citenamefont {Botzung}\ and\ \citenamefont {Nataf}(2023)}]{botzung2023exact}%
  \BibitemOpen
  \bibfield  {author} {\bibinfo {author} {\bibfnamefont {T.}~\bibnamefont {Botzung}}\ and\ \bibinfo {author} {\bibfnamefont {P.}~\bibnamefont {Nataf}},\ }\href@noop {} {\enquote {\bibinfo {title} {Exact diagonalization of $\mathrm{SU}(n)$ fermi-hubbard models},}\ } (\bibinfo {year} {2023}),\ \Eprint {http://arxiv.org/abs/2309.09965} {arXiv:2309.09965 [cond-mat.str-el]} \BibitemShut {NoStop}%
\bibitem [{\citenamefont {Nataf}\ and\ \citenamefont {Mila}(2014)}]{nataf_exact_2014}%
  \BibitemOpen
  \bibfield  {author} {\bibinfo {author} {\bibfnamefont {P.}~\bibnamefont {Nataf}}\ and\ \bibinfo {author} {\bibfnamefont {F.}~\bibnamefont {Mila}},\ }\href {https://link.aps.org/doi/10.1103/PhysRevLett.113.127204} {\bibfield  {journal} {\bibinfo  {journal} {Phys. Rev. Lett.}\ }\textbf {\bibinfo {volume} {113}},\ \bibinfo {pages} {127204} (\bibinfo {year} {2014})}\BibitemShut {NoStop}%
\bibitem [{\citenamefont {Wan}\ \emph {et~al.}(2017)\citenamefont {Wan}, \citenamefont {Nataf},\ and\ \citenamefont {Mila}}]{wan_exact_2017}%
  \BibitemOpen
  \bibfield  {author} {\bibinfo {author} {\bibfnamefont {K.}~\bibnamefont {Wan}}, \bibinfo {author} {\bibfnamefont {P.}~\bibnamefont {Nataf}}, \ and\ \bibinfo {author} {\bibfnamefont {F.}~\bibnamefont {Mila}},\ }\href {https://link.aps.org/doi/10.1103/PhysRevB.96.115159} {\bibfield  {journal} {\bibinfo  {journal} {Phys. Rev. B}\ }\textbf {\bibinfo {volume} {96}},\ \bibinfo {pages} {115159} (\bibinfo {year} {2017})}\BibitemShut {NoStop}%
\bibitem [{Note1()}]{Note1}%
  \BibitemOpen
  \bibinfo {note} {Or more precisely of its universal enveloping algebra since $H$ is a polynomial of the generators of the Lie Algebra of $\protect \mathrm {U}(L)$.}\BibitemShut {Stop}%
\bibitem [{\citenamefont {Weyl}(1925)}]{Weyl1925Dec}%
  \BibitemOpen
  \bibfield  {author} {\bibinfo {author} {\bibfnamefont {H.}~\bibnamefont {Weyl}},\ }\href {\doibase 10.1007/BF01506234} {\bibfield  {journal} {\bibinfo  {journal} {Math. Z.}\ }\textbf {\bibinfo {volume} {23}},\ \bibinfo {pages} {271} (\bibinfo {year} {1925})}\BibitemShut {NoStop}%
\bibitem [{\citenamefont {de~B.~Robinson}(1961)}]{Robinson1961}%
  \BibitemOpen
  \bibfield  {author} {\bibinfo {author} {\bibfnamefont {G.}~\bibnamefont {de~B.~Robinson}},\ }\href@noop {} {\emph {\bibinfo {title} {{Representation theory of the symmetric group}}}}\ (\bibinfo  {publisher} {Edinburgh: EUP},\ \bibinfo {year} {1961})\BibitemShut {NoStop}%
\bibitem [{\citenamefont {Paldus}(2021)}]{Paldus2021Jan}%
  \BibitemOpen
  \bibfield  {author} {\bibinfo {author} {\bibfnamefont {J.}~\bibnamefont {Paldus}},\ }\href {\doibase 10.1007/s10910-020-01174-7} {\bibfield  {journal} {\bibinfo  {journal} {J. Math. Chem.}\ }\textbf {\bibinfo {volume} {59}},\ \bibinfo {pages} {72} (\bibinfo {year} {2021})}\BibitemShut {NoStop}%
\bibitem [{\citenamefont {Gelfand}\ and\ \citenamefont {Tsetlin}(1950)}]{Gelfand_1950}%
  \BibitemOpen
  \bibfield  {author} {\bibinfo {author} {\bibfnamefont {I.~M.}\ \bibnamefont {Gelfand}}\ and\ \bibinfo {author} {\bibfnamefont {M.~L.}\ \bibnamefont {Tsetlin}},\ }\href@noop {} {\bibfield  {journal} {\bibinfo  {journal} {Dokl. Akad. Nauk SSSR}\ }\textbf {\bibinfo {volume} {71}},\ \bibinfo {pages} {825} (\bibinfo {year} {1950})}\BibitemShut {NoStop}%
\bibitem [{\citenamefont {Vilenkin}\ and\ \citenamefont {Klimyk}(1992)}]{Vilenkin_vol3}%
  \BibitemOpen
  \bibfield  {author} {\bibinfo {author} {\bibfnamefont {N.~J.}\ \bibnamefont {Vilenkin}}\ and\ \bibinfo {author} {\bibfnamefont {A.~U.}\ \bibnamefont {Klimyk}},\ }\href@noop {} {\emph {\bibinfo {title} {Representation of Lie groups and special functions, Vol. 3}}}\ (\bibinfo  {publisher} {Kluwer Academic Publishers},\ \bibinfo {year} {1992})\BibitemShut {NoStop}%
\bibitem [{\citenamefont {Pilch}\ and\ \citenamefont {Schellekens}(1984)}]{pilch_formulas_1984}%
  \BibitemOpen
  \bibfield  {author} {\bibinfo {author} {\bibfnamefont {K.}~\bibnamefont {Pilch}}\ and\ \bibinfo {author} {\bibfnamefont {A.~N.}\ \bibnamefont {Schellekens}},\ }\href {https://doi.org/10.1063/1.526101} {\bibfield  {journal} {\bibinfo  {journal} {Journal of Mathematical Physics}\ }\textbf {\bibinfo {volume} {25}},\ \bibinfo {pages} {3455} (\bibinfo {year} {1984})}\BibitemShut {NoStop}%
\bibitem [{Note2()}]{Note2}%
  \BibitemOpen
  \bibinfo {note} {When $M^A=L-1$ and $M^{\sigma }=0$ for $\sigma >1$: then only the fully symmetric (one row) irrep $\alpha $ is included in the considered subspace.}\BibitemShut {Stop}%
\bibitem [{Note3()}]{Note3}%
  \BibitemOpen
  \bibinfo {note} {Non separable means that it is still path connected if one site is withdrawn}\BibitemShut {NoStop}%
\bibitem [{\citenamefont {Xavier}\ \emph {et~al.}(2020)\citenamefont {Xavier}, \citenamefont {Kochetov},\ and\ \citenamefont {Ferraz}}]{Xavier_2020}%
  \BibitemOpen
  \bibfield  {author} {\bibinfo {author} {\bibfnamefont {H.~B.}\ \bibnamefont {Xavier}}, \bibinfo {author} {\bibfnamefont {E.}~\bibnamefont {Kochetov}}, \ and\ \bibinfo {author} {\bibfnamefont {A.}~\bibnamefont {Ferraz}},\ }\href {\doibase 10.1103/PhysRevB.101.045112} {\bibfield  {journal} {\bibinfo  {journal} {Phys. Rev. B}\ }\textbf {\bibinfo {volume} {101}},\ \bibinfo {pages} {045112} (\bibinfo {year} {2020})}\BibitemShut {NoStop}%
\bibitem [{\citenamefont {Yun}\ \emph {et~al.}(2021{\natexlab{b}})\citenamefont {Yun}, \citenamefont {Dobrautz}, \citenamefont {Luo},\ and\ \citenamefont {Alavi}}]{Sujun2021}%
  \BibitemOpen
  \bibfield  {author} {\bibinfo {author} {\bibfnamefont {S.}~\bibnamefont {Yun}}, \bibinfo {author} {\bibfnamefont {W.}~\bibnamefont {Dobrautz}}, \bibinfo {author} {\bibfnamefont {H.}~\bibnamefont {Luo}}, \ and\ \bibinfo {author} {\bibfnamefont {A.}~\bibnamefont {Alavi}},\ }\href {\doibase 10.1103/PhysRevB.104.235102} {\bibfield  {journal} {\bibinfo  {journal} {Phys. Rev. B}\ }\textbf {\bibinfo {volume} {104}},\ \bibinfo {pages} {235102} (\bibinfo {year} {2021}{\natexlab{b}})}\BibitemShut {NoStop}%
\bibitem [{\citenamefont {Lieb}\ and\ \citenamefont {Mattis}(1962{\natexlab{b}})}]{Lieb1962}%
  \BibitemOpen
  \bibfield  {author} {\bibinfo {author} {\bibfnamefont {E.}~\bibnamefont {Lieb}}\ and\ \bibinfo {author} {\bibfnamefont {D.}~\bibnamefont {Mattis}},\ }\href {\doibase 10.1103/PhysRev.125.164} {\bibfield  {journal} {\bibinfo  {journal} {Phys. Rev.}\ }\textbf {\bibinfo {volume} {125}},\ \bibinfo {pages} {164} (\bibinfo {year} {1962}{\natexlab{b}})}\BibitemShut {NoStop}%
\bibitem [{\citenamefont {White}(1992)}]{white_density_1992}%
  \BibitemOpen
  \bibfield  {author} {\bibinfo {author} {\bibfnamefont {S.~R.}\ \bibnamefont {White}},\ }\href {\doibase 10.1103/PhysRevLett.69.2863} {\bibfield  {journal} {\bibinfo  {journal} {Physical Review Letters}\ }\textbf {\bibinfo {volume} {69}},\ \bibinfo {pages} {2863} (\bibinfo {year} {1992})}\BibitemShut {NoStop}%
\bibitem [{\citenamefont {Singh}\ and\ \citenamefont {Oitmaa}(2022)}]{Singh_2022}%
  \BibitemOpen
  \bibfield  {author} {\bibinfo {author} {\bibfnamefont {R.~R.~P.}\ \bibnamefont {Singh}}\ and\ \bibinfo {author} {\bibfnamefont {J.}~\bibnamefont {Oitmaa}},\ }\href {\doibase 10.1103/PhysRevB.106.014424} {\bibfield  {journal} {\bibinfo  {journal} {Phys. Rev. B}\ }\textbf {\bibinfo {volume} {106}},\ \bibinfo {pages} {014424} (\bibinfo {year} {2022})}\BibitemShut {NoStop}%
\bibitem [{\citenamefont {Riera}\ and\ \citenamefont {Young}(1989)}]{Riera1989Sep}%
  \BibitemOpen
  \bibfield  {author} {\bibinfo {author} {\bibfnamefont {J.~A.}\ \bibnamefont {Riera}}\ and\ \bibinfo {author} {\bibfnamefont {A.~P.}\ \bibnamefont {Young}},\ }\href {\doibase 10.1103/PhysRevB.40.5285} {\bibfield  {journal} {\bibinfo  {journal} {Phys. Rev. B}\ }\textbf {\bibinfo {volume} {40}},\ \bibinfo {pages} {5285} (\bibinfo {year} {1989})}\BibitemShut {NoStop}%
\bibitem [{\citenamefont {Yun}\ \emph {et~al.}(2023)\citenamefont {Yun}, \citenamefont {Dobrautz}, \citenamefont {Luo}, \citenamefont {Katukuri}, \citenamefont {Liebermann},\ and\ \citenamefont {Alavi}}]{Yun2023Feb}%
  \BibitemOpen
  \bibfield  {author} {\bibinfo {author} {\bibfnamefont {S.}~\bibnamefont {Yun}}, \bibinfo {author} {\bibfnamefont {W.}~\bibnamefont {Dobrautz}}, \bibinfo {author} {\bibfnamefont {H.}~\bibnamefont {Luo}}, \bibinfo {author} {\bibfnamefont {V.}~\bibnamefont {Katukuri}}, \bibinfo {author} {\bibfnamefont {N.}~\bibnamefont {Liebermann}}, \ and\ \bibinfo {author} {\bibfnamefont {A.}~\bibnamefont {Alavi}},\ }\href {\doibase 10.1103/PhysRevB.107.064405} {\bibfield  {journal} {\bibinfo  {journal} {Phys. Rev. B}\ }\textbf {\bibinfo {volume} {107}},\ \bibinfo {pages} {064405} (\bibinfo {year} {2023})}\BibitemShut {NoStop}%
\bibitem [{\citenamefont {Weichselbaum}(2012)}]{weichelbaum_nonabelian_2012}%
  \BibitemOpen
  \bibfield  {author} {\bibinfo {author} {\bibfnamefont {A.}~\bibnamefont {Weichselbaum}},\ }\href {http://www.sciencedirect.com/science/article/pii/S0003491612001121} {\bibfield  {journal} {\bibinfo  {journal} {Annals of Physics}\ }\textbf {\bibinfo {volume} {327}},\ \bibinfo {pages} {2972 } (\bibinfo {year} {2012})}\BibitemShut {NoStop}%
\bibitem [{\citenamefont {Weichselbaum}\ \emph {et~al.}(2018)\citenamefont {Weichselbaum}, \citenamefont {Capponi}, \citenamefont {Lecheminant}, \citenamefont {Tsvelik},\ and\ \citenamefont {L\"auchli}}]{weichselbaum_unified_2018}%
  \BibitemOpen
  \bibfield  {author} {\bibinfo {author} {\bibfnamefont {A.}~\bibnamefont {Weichselbaum}}, \bibinfo {author} {\bibfnamefont {S.}~\bibnamefont {Capponi}}, \bibinfo {author} {\bibfnamefont {P.}~\bibnamefont {Lecheminant}}, \bibinfo {author} {\bibfnamefont {A.~M.}\ \bibnamefont {Tsvelik}}, \ and\ \bibinfo {author} {\bibfnamefont {A.~M.}\ \bibnamefont {L\"auchli}},\ }\href {https://link.aps.org/doi/10.1103/PhysRevB.98.085104} {\bibfield  {journal} {\bibinfo  {journal} {Phys. Rev. B}\ }\textbf {\bibinfo {volume} {98}},\ \bibinfo {pages} {085104} (\bibinfo {year} {2018})}\BibitemShut {NoStop}%
\bibitem [{\citenamefont {Nataf}\ and\ \citenamefont {Mila}(2018)}]{nataf_density_2018}%
  \BibitemOpen
  \bibfield  {author} {\bibinfo {author} {\bibfnamefont {P.}~\bibnamefont {Nataf}}\ and\ \bibinfo {author} {\bibfnamefont {F.}~\bibnamefont {Mila}},\ }\href {https://link.aps.org/doi/10.1103/PhysRevB.97.134420} {\bibfield  {journal} {\bibinfo  {journal} {Phys. Rev. B}\ }\textbf {\bibinfo {volume} {97}},\ \bibinfo {pages} {134420} (\bibinfo {year} {2018})}\BibitemShut {NoStop}%
\bibitem [{\citenamefont {Gozel}\ \emph {et~al.}(2020)\citenamefont {Gozel}, \citenamefont {Nataf},\ and\ \citenamefont {Mila}}]{gozel_2020}%
  \BibitemOpen
  \bibfield  {author} {\bibinfo {author} {\bibfnamefont {S.}~\bibnamefont {Gozel}}, \bibinfo {author} {\bibfnamefont {P.}~\bibnamefont {Nataf}}, \ and\ \bibinfo {author} {\bibfnamefont {F.}~\bibnamefont {Mila}},\ }\href {\doibase 10.1103/PhysRevLett.125.057202} {\bibfield  {journal} {\bibinfo  {journal} {Phys. Rev. Lett.}\ }\textbf {\bibinfo {volume} {125}},\ \bibinfo {pages} {057202} (\bibinfo {year} {2020})}\BibitemShut {NoStop}%
\end{thebibliography}%

\end{document}